\documentclass[graybox]{svmult}

\usepackage{amsmath,amssymb,bm}

\DeclareSymbolFont{varletters}{OML}{cmm}{m}{it}
\DeclareMathSymbol{\epsilon}{\mathalpha}{varletters}{"0F}
\DeclareSymbolFont{bfletters}{OML}{cmm}{b}{it}
\DeclareMathSymbol{\bfepsilon}{\mathalpha}{bfletters}{"0F}
\DeclareMathAlphabet{\mathcal}{OMS}{cmsy}{m}{n}

\newcommand\eps{\epsilon}
\renewcommand\d{\partial}
\newcommand\grad{\bm{\partial}}
\newcommand\+{\dagger}
\newcommand\<{\langle}
\renewcommand\>{\rangle}
\newcommand\p{{\bm{p}}}
\renewcommand\k{{\bm{k}}}
\newcommand\ep{\varepsilon_\p}
\newcommand\ek{\varepsilon_\k}
\newcommand\Ep{E_\p}
\newcommand\up{\uparrow}
\newcommand\down{\downarrow}
\newcommand\updown{{\up\down}}
\newcommand\kF{k_{\mathrm{F}}}
\newcommand\eF{\varepsilon_{\mathrm{F}}}
\newcommand\Veff{V_\mathrm{eff}}
\newcommand\Tc{{T_\mathrm{c}}}
\newcommand\x{{\bm{x}}}
\newcommand\y{{\bm{y}}}
\newcommand\0{{\bm{0}}}
\renewcommand\O{\mathcal{O}}

\usepackage{mathptmx}       
\usepackage{helvet}         
\usepackage{courier}        
\usepackage{type1cm}        
\usepackage{makeidx}         
\usepackage{graphicx}        
\usepackage{multicol}        
\usepackage[bottom]{footmisc}

\makeindex             

\begin{document}

\title*{Unitary Fermi gas, $\bfepsilon$ expansion,
and nonrelativistic conformal field theories}
\titlerunning{Unitary Fermi gas, $\epsilon$ expansion,
and nonrelativistic conformal field theories}

\author{Yusuke Nishida and Dam Thanh Son}

\institute{Yusuke Nishida \at Center for Theoretical Physics,
Massachusetts Institute of Technology \\
Cambridge, Massachusetts 02139, USA;
\ \email{nishida@mit.edu}
\and Dam Thanh Son \at Institute for Nuclear Theory,
University of Washington \\
Seattle, Washington 98195, USA;
\ \email{son@phys.washington.edu}}
%
%
\maketitle

\abstract{We review theoretical aspects of unitary Fermi gas (UFG),
which has been realized in ultracold atom experiments.  We first
introduce the $\epsilon$ expansion technique based on a systematic
expansion in terms of the dimensionality of space.  We apply this
technique to compute the thermodynamic quantities, the quasiparticle
spectrum, and the critical temperature of UFG.  We then discuss
consequences of the scale and conformal invariance of UFG.  We prove a
correspondence between primary operators in nonrelativistic conformal
field theories and energy eigenstates in a harmonic potential.  We use
this correspondence to compute energies of fermions at unitarity in a
harmonic potential.  The scale and conformal invariance together with
the general coordinate invariance constrains the properties of UFG.  We
show the vanishing bulk viscosities of UFG and derive the low-energy
effective Lagrangian for the superfluid UFG.  Finally we propose other
systems exhibiting the nonrelativistic scaling and conformal symmetries
that can be in principle realized in ultracold atom experiments.}

\newpage
\setcounter{minitocdepth}{5}
\dominitoc
\newpage

\section{Introduction}
Interacting fermions appear in various subfields of physics.  The
Bardeen-Cooper-Schrieffer (BCS) mechanism shows that if the interaction
is attractive, the Fermi surface is unstable toward the formation of
Cooper pairs and the ground state of the system exhibits
superconductivity or superfluidity.  Such phenomena have been observed
in metallic superconductors, superfluid $^3$He, and high-$\Tc$
superconductors.  Possibilities of superfluid nuclear matter, color
superconductivity of quarks, and neutrino superfluidity are also
discussed in literatures; some of these states might be important to the
physics of neutron stars.

In 2004, a new type of fermionic superfluid has been realized in
ultracold atomic gases of $^{40}$K and $^6$Li in optical
traps~\cite{Regal:2004,Zwierlein:2004}.  Unlike the previous examples,
these systems have a remarkable feature that the strength of the
attraction between fermions can be arbitrarily tuned through magnetic
field induced Feshbach resonances.  The interatomic interaction at
ultracold temperature is dominated by binary $s$-wave collisions, whose
strength is characterized by the $s$-wave scattering length $a$.  Across
the Feshbach resonance, $a^{-1}$ can, in principle, be tuned to any
value from $-\infty$ to $+\infty$.  Therefore, the ultracold atomic
gases provide an ideal ground for studying quantum physics of
interacting fermions from weak coupling to strong coupling.

In cold and dilute atomic gases, the interatomic potential is well
approximated by a zero-range contact interaction: the potential
range, $r_0\sim60\,a_0$ for $^{40}$K and $r_0\sim30\,a_0$ for $^6$Li, is
negligible compared to the de Broglie wavelength and the mean
interparticle distance ($n^{-1/3}\sim5000$-$10000\,a_0$).  The
properties of such a system are {\em universal\/}, i.e., independent of
details of the interaction potential.  By regarding the two different
hyperfine states of fermionic atoms as spin-$\up$ and spin-$\down$
fermions, the atomic gas reduces to a gas of spin-$\frac12$ fermions
interacting by the zero-range potential with the tunable scattering
length $a$.  The question we would like to understand is the phase
diagram of such a system as a function of the dimensionless parameter
$-\infty<(a\kF)^{-1}<\infty$, where $\kF\equiv(3\pi^2n)^{1/3}$ is the
Fermi momentum.

\begin{figure}[t]
 \begin{center}
  \includegraphics[width=0.95\textwidth]{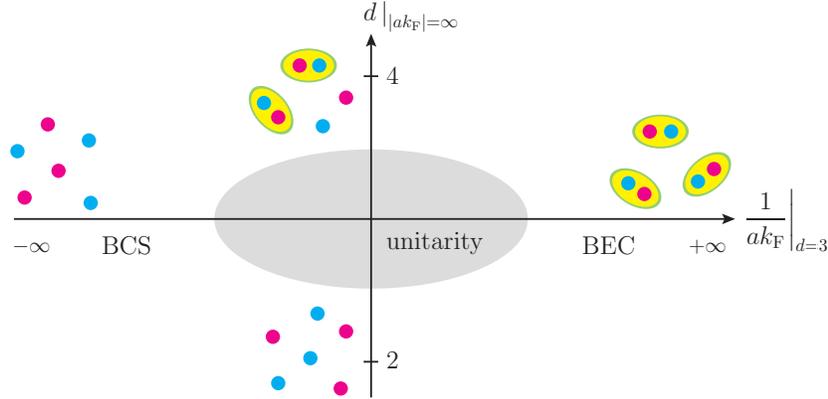}
 \end{center}
 \caption{Extended phase diagram of the BCS-BEC crossover in the plane
 of the inverse scattering length $(a\kF)^{-1}$ and the spatial
 dimension $d$.  There are four limits where the system becomes
 noninteracting; $a\kF\to\pm0$ and $d\to4,2$.  The system is strongly
 interacting in the shaded region. \label{fig:bcs-bec}}
\end{figure}

The qualitative understanding of the phase diagram is provided by picture
of a BCS-BEC crossover~\cite{Eagles:1969,Leggett:1980,Nozieres:1985}
(the horizontal axis in Fig.~\ref{fig:bcs-bec}).  When the attraction
between fermions is weak (BCS limit where $a\kF<0$ and $|a\kF|\ll1$),
the system is a weakly interacting Fermi gas.  Its ground state is
superfluid by the BCS mechanism, where (loosely bound) Cooper pairs
condense.  On the other hand, when the attraction is strong (BEC limit
where $0<a\kF\ll1$), two fermions form a bound molecule and the system
becomes a weakly interacting Bose gas of such molecules.  Its ground
state again exhibits superfluidity, but by the Bose-Einstein
condensation (BEC) of the tightly bound molecules.  These two regimes
are smoothly connected without phase transitions, which implies that the
ground state of the system is a superfluid for any $(a\kF)^{-1}$.  Both
BCS and BEC limits can be understood quantitatively by using the
standard perturbative expansion in terms of the small parameter
$|a\kF|\ll1$.

In contrast, a strongly interacting regime exists in the middle of the
BCS-BEC crossover, where the scattering length is comparable to or
exceeds the mean interparticle distance; $|a\kF|\gtrsim1$.  In
particular, the limit of infinite scattering length $|a\kF|\to\infty$,
which is often called the {\em unitarity\/} limit, has attracted intense
attention by experimentalists and theorists alike.  Beside being
experimentally realizable in ultracold atomic gases using the Feshbach
resonance, this regime is an idealization of the dilute
nuclear matter, where the neutron-neutron scattering length
$a_{nn}\simeq-18.5\ \mathrm{fm}$ is larger than the typical range of the
nuclear force $r_0\simeq1.4\ \mathrm{fm}$.

Theoretical treatments of the Fermi gas in the unitarity limit (unitary
Fermi gas) suffer from the difficulty arising from the lack of a small
expansion parameter: the standard perturbative expansion in terms of
$|a\kF|$ is obviously of no use.  Mean-field type approximations, with
or without fluctuations, are often adopted to obtain a qualitative
understanding of the BCS-BEC crossover, but they are not necessarily
controlled near the unitarity limit.  Therefore an important and
challenging problem for theorists is to establish a {\em systematic\/}
approach to investigate the unitary Fermi gas.

In Sect.~\ref{sec:epsilon} of this chapter we describe one of such
approaches.  It is based on an expansion over a parameter which depends
on the dimensionality of
space~\cite{Nishida:2006br,Nishida:2006eu,Nishida:2006rp}.  In this
approach, one extends the problem to arbitrary spatial dimension $d$,
keeping the scattering length infinite $|a\kF|\to\infty$ (the vertical
axis in Fig.~\ref{fig:bcs-bec}).  Then we can find two noninteracting
limits on the $d$ axis, which are $d=4$ and $d=2$.  Accordingly,
slightly below four or slightly above two spatial dimensions, the
unitary Fermi gas becomes weakly interacting and thus a ``perturbative
expansion'' is available.  We show that the unitary Fermi gas near $d=4$
is described by a weakly interacting gas of bosons and fermions, while
near $d=2$ it reduces to a weakly interacting Fermi gas.  A small
parameter for the perturbative expansion is $\eps\equiv4-d$ near four
spatial dimensions or $\bar\eps\equiv d-2$ near two spatial dimensions.
After performing all calculations treating $\eps$ or $\bar\eps$ as a
small expansion parameter, results for the physical case of $d=3$ are
obtained by extrapolating the series expansions to $\eps\,(\bar\eps)=1$,
or more appropriately, by matching the two series expansions.  We apply
this technique, the $\eps$ expansion, to compute the thermodynamic
quantities~\cite{Nishida:2006br,Nishida:2006eu,Arnold:2006fr,Nishida:2008mh}
(Sect.~\ref{sec:zero_T}), the quasiparticle
spectrum~\cite{Nishida:2006br,Nishida:2006eu}
(Sect.~\ref{sec:spectrum}), and the critical
temperature~\cite{Nishida:2006rp} (Sect.~\ref{sec:Tc}).  The main
advantage of this method is that all calculations can be done
analytically; its drawback is that interpolations to $d=3$ are needed to
achieve numerical accuracy.

Then in Sects.~\ref{sec:NRCFT} and \ref{sec:general}, we focus on
consequences of another important characteristic of the unitary Fermi
gas, namely, the scale and conformal
invariance~\cite{Nishida:2007pj,Son:2005tj,Son:2005rv}.  We introduce
the notion of nonrelativistic conformal field theories (NRCFTs) as
theories describing nonrelativistic systems exhibiting the scaling and
conformal symmetries.  In Sect.~\ref{sec:algebra}, we describe a
nonrelativistic analog of the conformal algebra, the so-called
Schr\"odinger algebra~\cite{Hagen:1972pd,Niederer:1972zz}, and show in
Sect.~\ref{sec:correspondence} that there is an operator-state
correspondence~\cite{Nishida:2007pj}:  a primary operator in NRCFT
corresponds to an energy eigenstate of a few-particle system in a
harmonic potential.  The scaling dimension of the primary operator
coincides with the energy eigenvalue of the corresponding state, divided
by the oscillator frequency.  We use the operator-state correspondence
to compute the energies of two and three fermions at unitarity in a
harmonic potential exactly (Sect.~\ref{sec:scaling} with Appendix) and
more fermions with the help of the $\eps$ expansion
(Sect.~\ref{sec:application}).

The enlarged symmetries of the unitary Fermi gas also constrain its
properties.  By requiring the scale and conformal invariance and the
general coordinate invariance of the hydrodynamic equations, we show
that the unitary Fermi gas has the vanishing bulk viscosity in the
normal phase~\cite{Son:2005tj} (Sect.~\ref{sec:viscosity}).  In the
superfluid phase, two of the three bulk viscosities have to vanish while
the third one is allowed to be nonzero.  In Sect.~\ref{sec:EFT}, we
derive the most general effective Lagrangian for the superfluid unitary
Fermi gas that is consistent with the scale, conformal, and general
coordinate invariance in the systematic momentum
expansion~\cite{Son:2005rv}.  To the leading and next-to-leading orders,
there are three low-energy constants which can be computed using the
$\eps$ expansion.  We can express various physical quantities through
these constants.

Finally in Sect.~\ref{sec:other}, we discuss other systems exhibiting
the nonrelativistic scaling and conformal symmetries, to which a part of
above results can be applied.  Such systems include a mass-imbalanced
Fermi gas with both two-body and three-body
resonances~\cite{Nishida:2007mr} and Fermi gases in mixed
dimensions~\cite{Nishida:2008kr}.  These systems can be in principle
realized in ultracold atom experiments.

\section[$\epsilon$ expansion for the unitary Fermi gas]{$\bfepsilon$ expansion for the unitary Fermi gas
 \label{sec:epsilon}}
In this section, we develop an analytical approach for the unitary Fermi
gas based on a systematic expansion in terms of the dimensionality of
space by using special features of four or two spatial dimensions for
the zero-range and infinite scattering length interaction.

\subsection{Why four and two spatial dimensions are special?}

\subsubsection{Nussinovs' intuitive arguments \label{sec:nussinov}}
The special role of four and two spatial dimensions for the zero-range
and infinite scattering length interaction has been first recognized by
Nussinov and Nussinov~\cite{Nussinov:2006}.  At infinite scattering
length, which corresponds to a resonance at zero energy, the two-body
wave function at a short distance $r\to0$ behaves like
\begin{equation}\label{eq:wave_function}
 R(r) \propto \frac1{r^{d-2}} + O(r^{4-d}),
\end{equation}
where $r$ is the separation between two fermions with opposite spins.
The first singular term $\sim1/r^{d-2}$ is the spherically symmetric
solution to the Laplace equation in $d$ spatial dimensions.  Accordingly
the normalization integral of the wave function has the form
\begin{equation}
 \int\!d\bm{r}R(r)^2 \propto \int_0dr\frac{1}{r^{d-3}},
\end{equation}
which diverges at the origin $r\to0$ in higher dimensions $d\geq4$.
Therefore, in the limit $d\to4$, the two-body wave function is
concentrated at the origin and the fermion pair should behave like a
point-like composite boson.  This observation led Nussinov and Nussinov
to conclude that the unitary Fermi gas at $d\to4$ becomes a
noninteracting Bose gas.

On the other hand, the singularity in the wave function
(\ref{eq:wave_function}) disappears in the limit $d\to2$, which means
that the interaction between the two fermions also disappears.  This can
be understood intuitively from the fact that in lower dimensions
$d\leq2$, any attractive potential possesses at least one bound state
and thus the threshold of the appearance of the first bound state
(infinite scattering length) corresponds to the vanishing potential.
Therefore the unitary Fermi gas at $d\to2$ should reduce to a
noninteracting Fermi gas~\cite{Nussinov:2006}.

The physical case, $d=3$, lies midway between these two limits $d=2$ and
$d=4$.  It seems natural to try to develop an expansion around these two
limits and then extrapolate to $d=3$.  For this purpose, we need to
employ a field theoretical approach.

\subsubsection{Field theoretical approach \label{sec:field_theory}}
Spin-$\frac12$ fermions interacting by the zero-range potential is
described by the following Lagrangian density (here and below
$\hbar=k_\mathrm{B}=1$):
\begin{equation}\label{eq:L}
 \mathcal{L} = \sum_{\sigma=\up,\down}\psi_\sigma^\+
  \left(i\d_t+\frac{\grad^2}{2m}+\mu\right)\psi_\sigma
  +c_0\psi_\up^\+\psi_\down^\+\psi_\down\psi_\up.
\end{equation}
In $d=3$, the bare coupling $c_0$ is related to the physical parameter,
the scattering length $a$, by
\begin{equation}
 \frac1{c_0} = -\frac{m}{4\pi a} + \int\!\frac{d\k}{(2\pi)^d}\frac1{2\ek}
  \qquad \left(\ek\equiv\frac{\k^2}{2m}\right).
\end{equation}
In dimensional regularization, used in this section, the second term
vanishes and therefore the unitarity limit $a\to\infty$ corresponds to
$c_0\to\infty$.

\begin{figure}[t]
 \begin{center}
  \includegraphics[width=\textwidth]{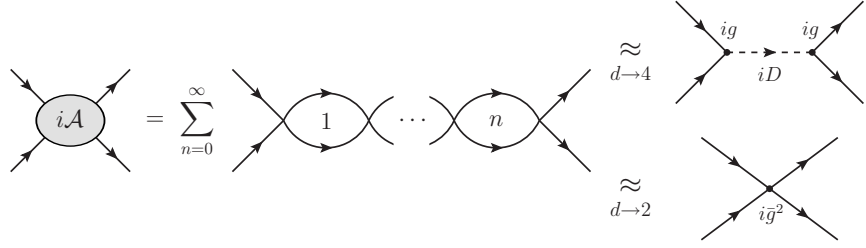}
 \end{center}
 \caption{Two-fermion scattering in vacuum in the unitarity limit. The
 scattering amplitude $i\mathcal{A}$ near $d=4$ is expressed by the
 propagation of a boson with the small effective coupling $g$, while it
 reduces to a contact interaction with the small effective coupling
 $\bar g^2$ near $d=2$. \label{fig:scattering}}
\end{figure}

In order to understand the specialty of $d=4$ and $d=2$, we first study
the scattering of two fermions in vacuum ($\mu=0$) in general $d$
spatial dimensions.  The two-body scattering amplitude
$i\mathcal{A}$ is given by the geometric series of bubble diagrams
depicted in Fig.~\ref{fig:scattering}.  In the unitarity limit, we
obtain
\begin{equation}
 i\mathcal{A}(p_0,\p) = -i\frac{\left(\frac{4\pi}m\right)^{d/2}}
  {\Gamma\!\left(1-\frac{d}2\right)\left(-p_0+\frac{\ep}2-i0^+\right)^{d/2-1}},
\end{equation}
which vanishes when $d\to4$ and $d\to2$ because of the poles in
$\Gamma\!\left(1-\frac{d}2\right)$.  This means that those dimensions
correspond to the noninteracting limits and is consistent with the
Nussinovs' arguments in Sect.~\ref{sec:nussinov}.

Furthermore, by expanding $i\mathcal{A}$ in terms of $\eps=4-d\ll1$, we
can obtain further insight.  The scattering amplitude to the leading
order in $\eps$ becomes
\begin{eqnarray}\label{eq:T_4d}
 i\mathcal{A}(p_0,\p) &=& \nonumber
  -\frac{8\pi^2\eps}{m^2}\frac{i}{p_0-\frac{\ep}2+i0^+} + O(\eps^2) \\
  &\equiv& (ig)^2iD(p_0,\p) + O(\eps^2),
\end{eqnarray}
where we have defined $g^2=\frac{8\pi^2\eps}{m^2}$ and
$D(p_0,\p)=\left(p_0-\frac{\ep}2+i0^+\right)^{-1}$.  The latter is the
propagator of a particle of mass $2m$.  This particle is a boson, which
is the point-like composite of two fermions.  Equation~(\ref{eq:T_4d})
states that the two-fermion scattering near $d=4$ can be thought of as
occurring through the propagation of an intermediate boson, as depicted
in Fig.~\ref{fig:scattering}.  The effective coupling of the two
fermions with the boson is $g\sim\eps^{1/2}$, which becomes small near
$d=4$.  This indicates the possibility to formulate a systematic
perturbative expansion for the unitary Fermi gas around $d=4$ as a
weakly interacting fermions and bosons.

Similarly, by expanding $i\mathcal{A}$ in $\bar\eps=d-2\ll1$, the
scattering amplitude becomes
\begin{equation}\label{eq:T_2d}
 i\mathcal{A}(p_0,\p) = i\frac{2\pi}{m}\bar\eps + O(\bar\eps^2)
  \equiv i\bar g^2 + O(\bar\eps^2),
\end{equation}
where we have defined $\bar g^2=\frac{2\pi}{m}\bar\eps$.
Equation~(\ref{eq:T_2d}) shows that the two-fermion scattering near
$d=2$ reduces to that caused by a contact interaction with the effective
coupling $\bar g^2$ (Fig.~\ref{fig:scattering}).  Because
$\bar g^2\sim\bar\eps$ is small near $d=2$, it will be possible to
formulate another systematic perturbative expansion for the unitary
Fermi gas around $d=2$ as a weakly interacting fermions.

\subsection[Feynman rules and power counting of $\epsilon$]{Feynman rules and power counting of $\bfepsilon$ \label{sec:formulation}}
The observations in Sect.~\ref{sec:field_theory} reveal how we should
construct the systematic expansions for the unitary Fermi gas around
$d=4$ and $d=2$.  Here we provide their formulations and power counting
rules of $\eps$ ($\bar\eps$).  The detailed derivations of the power
counting rules can be found in Ref.~\cite{Nishida:2006eu}.

\subsubsection{Around four spatial dimensions}
In order to organize a systematic expansion around $d=4$, we make a
Hubbard-Stratonovich transformation and rewrite the Lagrangian density
(\ref{eq:L}) as
\begin{subequations}
 \begin{eqnarray}
  \label{eq:Hubbard-Stratonovich1}
  \mathcal{L} &\to& \sum_{\sigma=\up,\down}\psi_\sigma^\+
   \left(i\d_t+\frac{\grad^2}{2m}+\mu\right)\psi_\sigma
   - \frac1{c_0}\phi^\+\phi + \psi_\up^\+\psi_\down^\+\phi
   + \phi^\+\psi_\down\psi_\up \\
  &=& \Psi^\+\left(i\d_t+\sigma_3\frac{\grad^2}{2m}+\sigma_3\mu\right)\Psi
   - \frac1{c_0}\phi^\+\phi + \Psi^\+\sigma_+\Psi\phi
   + \phi^\+\Psi^\+\sigma_-\Psi,
   \label{eq:Hubbard-Stratonovich2}
 \end{eqnarray}
\end{subequations}
where $\Psi\equiv(\psi_\up,\psi_\down^\+)^T$ is a two-component
Nambu-Gor'kov field, $\sigma_\pm\equiv\frac12(\sigma_1\pm i\sigma_2)$,
and $\sigma_{1,2,3}$ are the Pauli matrices.  Because the ground state
of the system at finite density is a superfluid, we expand $\phi$ around
its vacuum expectation value $\phi_0\equiv\<\phi\>>0$ as
\begin{equation}\label{eq:g_4d}
 \phi = \phi_0 + g\varphi
  \qquad\text{with}\qquad
  g \equiv \frac{\left(8\pi^2\eps\right)^{1/2}}{m}
  \left(\frac{m\phi_0}{2\pi}\right)^{\eps/4}.
\end{equation}
Here we introduced the effective coupling $g$ and chose the extra factor
$\left(m\phi_0/2\pi\right)^{\eps/4}$ so that $\varphi$ has the same
dimension as a nonrelativistic field.\footnote{The choice of the extra
factor is arbitrary, if it has the correct dimension, and does not
affect final results because the difference can be absorbed by the
redefinition of $\varphi$.  The particular choice of $g$ in
Eq.~(\ref{eq:g_4d}) [or $\bar g$ in Eq.~(\ref{eq:g_2d})] simplifies
expressions for loop integrals in the intermediate steps.}

\begin{figure}[t]
 \begin{center}
  \includegraphics[width=0.7\textwidth]{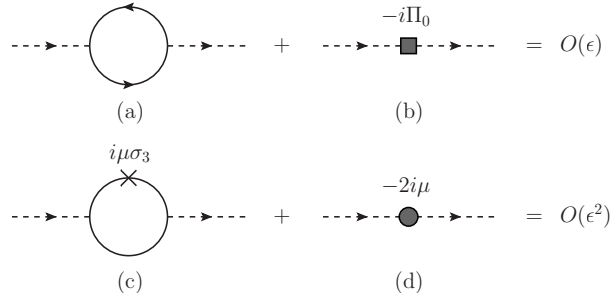}
 \end{center}
 \caption{Power counting rule of $\eps$.  The $1/\eps$ singularity in
 the boson self-energy diagram (a) or (c) can be canceled by combining
 it with the vertex from $\mathcal{L}_2$, (b) or (d), to achieve the
 simple $\eps$ counting.  Solid (dotted) lines are the fermion (boson)
 propagators $iG$ ($iD$).  The fermion loop in (c) goes around clockwise
 and counterclockwise and the cross symbol represents the $\mu$
 insertion. \label{fig:cancellation_4d}}
\end{figure}

Because the Lagrangian density (\ref{eq:Hubbard-Stratonovich2}) does not
have the kinetic term for the boson field $\varphi$, we add and subtract
its kinetic term by hand and rewrite the Lagrangian density as a sum of
three parts, $\mathcal{L}=\mathcal{L}_0+\mathcal{L}_1+\mathcal{L}_2$,
where
\begin{subequations}\label{eq:L_4d}
 \begin{eqnarray}
  \mathcal{L}_0 &=&
   \Psi^\+\left(i\d_t+\sigma_3\frac{\grad^2}{2m}+\sigma_1\phi_0\right)\Psi
   + \varphi^\+\left(i\d_t+\frac{\grad^2}{4m}\right)\varphi, \\
  \mathcal{L}_1 &=& g\Psi^\+\sigma_+\Psi\varphi 
   + g\varphi^\+\Psi^\+\sigma_-\Psi + \mu\Psi^\+\sigma_3\Psi
   + 2\mu\varphi^\+\varphi, \\
  \mathcal{L}_2 &=& -\varphi^\+\left(i\d_t+\frac{\grad^2}{4m}\right)\varphi
   - 2\mu\varphi^\+\varphi.
 \end{eqnarray}
\end{subequations}
Here we set $1/c_0=0$ in the unitarity limit.  We treat $\mathcal{L}_0$
as the unperturbed part and $\mathcal{L}_1$ as a small perturbation.
Note that the chemical potential $\mu$ is also treated as a perturbation
because we will find it small, $\mu/\phi_0\sim\eps$, by solving the gap
equation.  Physically, $\mathcal{L}_0+\mathcal{L}_1$ is the Lagrangian
density describing weakly interacting fermions $\Psi$ and bosons
$\varphi$ with a small coupling $g\sim\eps^{1/2}$.  $\mathcal{L}_2$
plays a role of counter terms that cancel $1/\eps$ singularities of loop
integrals in certain types of diagrams
(Fig.~\ref{fig:cancellation_4d}).

The unperturbed part $\mathcal{L}_0$ generates the fermion propagator,
\begin{equation}\label{eq:G_4d}
 G(p_0,\p) = \frac1{p_0^2-E_\p^2+i0^+}
  \begin{pmatrix}
   p_0+\ep & -\phi_0 \\
   -\phi_0 & p_0-\ep
  \end{pmatrix},
\end{equation}
with $E_\p\equiv\sqrt{\ep^2+\phi_0^2}$ and the boson propagator
\begin{equation}
 D(p_0,\p) = \frac1{p_0-\frac{\ep}2+i0^+}.
\end{equation}
The first two terms in the perturbation part $\mathcal{L}_1$ generate
the fermion-boson vertices whose coupling is $g$ in
Eq.~(\ref{eq:g_4d}).  The third and fourth terms are the chemical
potential insertions to the fermion and boson propagators. The two terms
in $\mathcal{L}_2$ provide additional vertices, $-i\Pi_0$ and $-2i\mu$
in Fig.~\ref{fig:cancellation_4d}(b) and (d), to the boson propagator,
where $\Pi_0(p_0,\p)\equiv p_0-\frac{\ep}2$.

The power counting rule of $\eps$ is simple and summarized as follows:
\begin{enumerate}
 \item We regard $\phi_0$ as $O(1)$ and hence $\mu\sim\eps\phi_0$ as
       $O(\eps)$.
 \item We write down Feynman diagrams for the quantity of interest using
       the propagators from $\mathcal{L}_0$ and the vertices from
       $\mathcal{L}_1$.
 \item If the written Feynman diagram includes any subdiagram of the
       type in Fig.~\ref{fig:cancellation_4d}(a) or
       Fig.~\ref{fig:cancellation_4d}(c), we add the same Feynman
       diagram where the subdiagram is replaced by the vertex from
       $\mathcal{L}_2$ in Fig.~\ref{fig:cancellation_4d}(b) or
       Fig.~\ref{fig:cancellation_4d}(d).
 \item The power of $\eps$ for the given Feynman diagram is
       $O(\eps^{N_g/2+N_\mu})$, where $N_g$ is the number of couplings
       $g$ and $N_\mu$ is the number of chemical potential insertions.
 \item The only exception is the one-loop vacuum diagram with one $\mu$
       insertion (the second diagram in Fig.~\ref{fig:vacuum_4d}), which
       is $O(1)$ instead of $O(\eps)$ due to the $1/\eps$ singularity
       arising from the loop integral.
\end{enumerate}
The same or similar power counting rule can be derived for the cases
with unequal chemical potentials
$\mu_\up\neq\mu_\down$~\cite{Nishida:2006eu}, unequal masses 
$m_\up\neq m_\down$~\cite{Nishida:2006wk}, at finite temperature
$T\neq0$~\cite{Nishida:2006rp}, and in the vicinity of the unitarity
limit $c_0\neq0$~\cite{Nishida:2006eu}.

\subsubsection{Around two spatial dimensions}
The systematic expansion around $d=2$ in terms of $\bar\eps$ can be also
organized in a similar way.  Starting with the Lagrangian density
(\ref{eq:Hubbard-Stratonovich2}), we expand $\phi$ around its vacuum
expectation value $\bar\phi_0\equiv\<\phi\>>0$ as
\begin{equation}\label{eq:g_2d}
 \phi = \bar\phi_0 + \bar g\varphi
  \qquad\text{with}\qquad
  \bar g \equiv \left(\frac{2\pi\bar\eps}m\right)^{1/2}
  \left(\frac{m\mu}{2\pi}\right)^{-\bar\eps/4}.
\end{equation}
Here we introduced the effective coupling $\bar g$ and chose the extra
factor $\left(m\mu/2\pi\right)^{-\bar\eps/4}$ so that
$\varphi^\+\varphi$ has the same dimension as the Lagrangian
density.\footnote{See the footnote after Eq.~(\ref{eq:g_4d}).}

\begin{figure}[t]
 \begin{center}
  \includegraphics[width=0.7\textwidth]{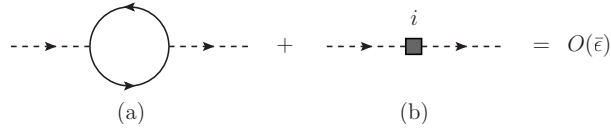}
 \end{center}
 \caption{Power counting rule of $\bar\eps$.  The $1/\bar\eps$
 singularity in the self-energy diagram of $\varphi$ field (a) can be
 canceled by combining it with the vertex from $\bar{\mathcal{L}}_2$ (b)
 to achieve the simple $\bar\eps$ counting.  Solid (dotted) lines are
 the fermion (auxiliary field) propagators $i\bar G$ ($i\bar D$).
 \label{fig:cancellation_2d}}
\end{figure}

In the unitarity limit $1/c_0=0$, we rewrite the Lagrangian density as a
sum of three parts,
$\mathcal{L}=\bar{\mathcal{L}}_0+\bar{\mathcal{L}}_1+\bar{\mathcal{L}}_2$,
where
\begin{subequations}\label{eq:L_2d}
 \begin{eqnarray}
  \bar{\mathcal{L}}_0 &=& \Psi^\+\left(i\d_t+\sigma_3\frac{\grad^2}{2m}
   +\sigma_3\mu+\sigma_1\bar\phi_0\right)\Psi, \\
  \bar{\mathcal{L}}_1 &=& -\varphi^\+\varphi
   + \bar g\Psi^\+\sigma_+\Psi\varphi
   + \bar g\varphi^\+\Psi^\+\sigma_-\Psi, \\
  \bar{\mathcal{L}}_2 &=& \varphi^\+\varphi.
 \end{eqnarray}
\end{subequations}
We treat $\bar{\mathcal{L}}_0$ as the unperturbed part and
$\bar{\mathcal{L}}_1$ as a small perturbation.  Physically,
$\bar{\mathcal{L}}_0+\bar{\mathcal{L}}_1$ is the Lagrangian density
describing weakly interacting fermions $\Psi$.  Indeed, if we did not
have $\bar{\mathcal{L}}_2$, we could integrate out the auxiliary fields
$\varphi$ and $\varphi^\+$,
\begin{equation}
 \bar{\mathcal{L}}_1 \to \bar g^2\Psi^\+\sigma_+\Psi\Psi^\+\sigma_-\Psi
  = \bar g^2\psi_\up^\+\psi_\down^\+\psi_\down\psi_\up,
\end{equation}
which is the contact interaction between fermions with a small coupling
$\bar g^2\sim\bar\eps$.  $\bar{\mathcal{L}}_2$ plays a role of a counter
term that cancels $1/\bar\eps$ singularities of loop integrals in a
certain type of diagrams (Fig.~\ref{fig:cancellation_2d}).

The unperturbed part $\bar{\mathcal{L}}_0$ generates the fermion
propagator
\begin{equation}
  \bar G(p_0,\p) = \frac1{p_0^2-\bar E_\p^2+i0^+}
  \begin{pmatrix}
   p_0+\ep-\mu & -\bar\phi_0 \\
   -\bar\phi_0 & p_0-\ep+\mu
  \end{pmatrix}
\end{equation}
with $\bar E_\p\equiv\sqrt{(\ep-\mu)^2+\bar\phi_0^2}$.  The first term in
the perturbation part $\bar{\mathcal{L}}_1$ generates the propagator of
$\varphi$ field, $\bar D(p_0,\p) = -1$, and the last two terms generate
the vertices between fermions and $\varphi$ field with the coupling
$\bar g$ in Eq.~(\ref{eq:g_2d}).  $\bar{\mathcal{L}}_2$ provides an
additional vertex $i$ in Fig.~\ref{fig:cancellation_2d}(b) to the
$\varphi$ propagator.

The power counting rule of $\bar\eps$ is simple and summarized as
follows:
\begin{enumerate}
 \item We regard $\mu$ as $O(1)$.
 \item We write down Feynman diagrams for the quantity of interest using
       the propagator from $\bar{\mathcal{L}}_0$ and the vertices from
       $\bar{\mathcal{L}}_1$.
 \item If the written Feynman diagram includes any subdiagram of the
       type in Fig.~\ref{fig:cancellation_2d}(a), we add the same
       Feynman diagram where the subdiagram is replaced by the vertex
       from $\bar{\mathcal{L}}_2$ in Fig.~\ref{fig:cancellation_2d}(b).
 \item The power of $\bar\eps$ for the given Feynman diagram is
       $O(\bar\eps^{N_{\bar g}/2})$, where $N_{\bar g}$ is the number of
       couplings $\bar g$.
\end{enumerate}

\subsection{Zero temperature thermodynamics \label{sec:zero_T}}
We now apply the developed $\eps$ expansion to compute various physical
quantities of the unitary Fermi gas.  Because of the absence of scales
in the zero-range and infinite scattering interaction, the density $n$
is the only scale of the unitary Fermi gas at zero temperature.
Therefore all physical quantities are determined by simple dimensional
analysis up to dimensionless constants of proportionality.  Such
dimensionless parameters are universal depending only on the
dimensionality of space.

A representative example of the universal parameters is the ground state
energy of the unitary Fermi gas normalized by that of a noninteracting
Fermi gas with the same density:
\begin{equation}
 \xi \equiv \frac{E_\mathrm{unitary}}{E_\mathrm{free}}.
\end{equation}
$\xi$, sometimes called the Bertsch parameter, measures how much energy
is gained due to the attractive interaction in the unitarity limit.
In terms of $\xi$, the pressure $P$, the energy density $\mathcal{E}$,
the chemical potential $\mu$, and the sound velocity $v_\mathrm{s}$ of
the unitary Fermi gas are given by
\begin{equation}\label{eq:pressure}
 \frac{P}{\eF n} = \frac{2}{d+2}\xi \qquad
  \frac{\mathcal{E}}{\eF n} = \frac{d}{d+2}\xi \qquad
  \frac{\mu}{\eF} = \xi \qquad
  \frac{v_\mathrm{s}}{v_\mathrm{F}} = \sqrt{\frac{\xi}{d}},
\end{equation}
where $\eF\equiv\kF^2/2m$ and $v_\mathrm{F}\equiv\kF/m$ are the Fermi
energy and velocity with
$\kF\equiv\left[2^{d-1}\pi^{d/2}\Gamma\!\left(\frac{d}2+1\right)n\right]^{1/d}$
being the Fermi momentum in $d$ spatial dimensions.  $\xi$ is thus a
fundamental quantity characterizing the zero temperature thermodynamics
of the unitary Fermi gas.

\subsubsection{Next-to-leading orders \label{sec:NLO}}

\begin{figure}[t]
 \begin{center}
  \includegraphics[width=0.75\textwidth]{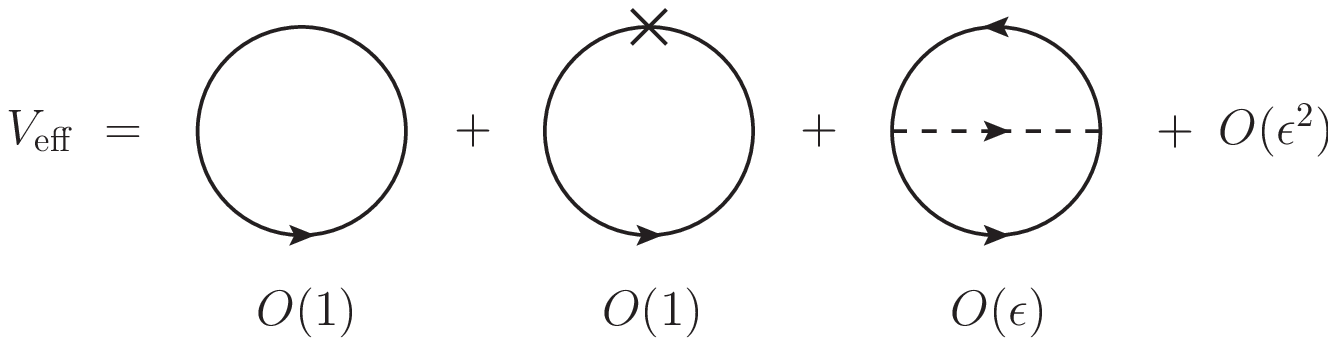}
 \end{center}
 \caption{Vacuum diagrams contributing to the effective potential near
 $d=4$ to leading and next-to-leading orders in $\eps$.  The boson
 one-loop diagram vanishes at zero temperature.  \label{fig:vacuum_4d}}
\end{figure}

$\xi$ can be determined systematically in the $\eps$ expansion by
computing the effective potential $\Veff(\phi_0)$, whose minimum with
respect to the order parameter $\phi_0$ provides the pressure
$P=-\min\Veff(\phi_0)$.  To leading and next-to-leading orders in
$\eps$, the effective potential receives contributions from three vacuum
diagrams depicted in Fig.~\ref{fig:vacuum_4d}.  The third diagram,
fermion loop with one boson exchange, results from the summation of
fluctuations around the classical solution and is beyond the mean field
approximation.  Any other diagrams are suppressed near $d=4$ by further
powers of $\eps$.  Performing the loop integrations, we obtain
\begin{eqnarray}
 \Veff(\phi_0) &=& \left[\frac{\phi_0}3
 \left\{1+\frac{7-3(\gamma+\ln2)}6\eps-3C\eps\right\}\right. \nonumber\\
 && - \left.\frac{\mu}{\eps}\left\{1+\frac{1-2(\gamma-\ln2)}4\eps\right\}\right]
 \left(\frac{m\phi_0}{2\pi}\right)^{d/2} + O(\eps^2),
\end{eqnarray}
where $C\approx0.14424$ is a numerical constant.  The minimization of
$\Veff(\phi_0)$ with respect to $\phi_0$ gives the gap equation,
$\d\Veff/\d\phi_0=0$, which is solved by
\begin{equation}\label{eq:phi_4d}
 \phi_0(\mu) = \frac{2\mu}\eps
  \left[1 + \left(3C-1+\ln2\right)\eps + O(\eps^2)\right].
\end{equation}
The effective potential at its minimum provides the pressure
$P(\mu)=-\Veff(\phi_0(\mu))$ as a function of $\mu$.  From
Eq.~(\ref{eq:pressure}) and $n=\d P/\d\mu$, we obtain $\xi$ expanded in
terms of $\eps$:
\begin{eqnarray}\label{eq:xi_4d}
 \left.\xi\right|_{d\to4} &=& \frac{\eps^{3/2}}2 \nonumber
  \exp\!\left(\frac{\eps\ln\eps}{8-2\eps}\right) 
  \left[1 - \left\{3C-\frac54(1-\ln2)\right\}\eps + O(\eps^2)\right] \\
 &=& \frac12\eps^{3/2} + \frac1{16}\eps^{5/2}\ln\eps
  - 0.0246\,\eps^{5/2} + O(\eps^{7/2}).
\end{eqnarray}

Although our formalism is based on the smallness of $\eps$, we find that
the next-to-leading-order correction is quite small compared to the
leading term even at $\eps=1$.  The naive extrapolation to the physical
dimension $d=3$ gives
\begin{equation}
 \xi \to 0.475 \qquad (\eps\to1),
\end{equation}
which is already close to the result from the Monte Carlo simulation
$\xi\approx0.40(1)$~\cite{Zhang}.  For comparison, the mean
field approximation yields $\xi\approx0.591$.

\begin{figure}[t]
 \begin{center}
  \includegraphics[width=0.55\textwidth]{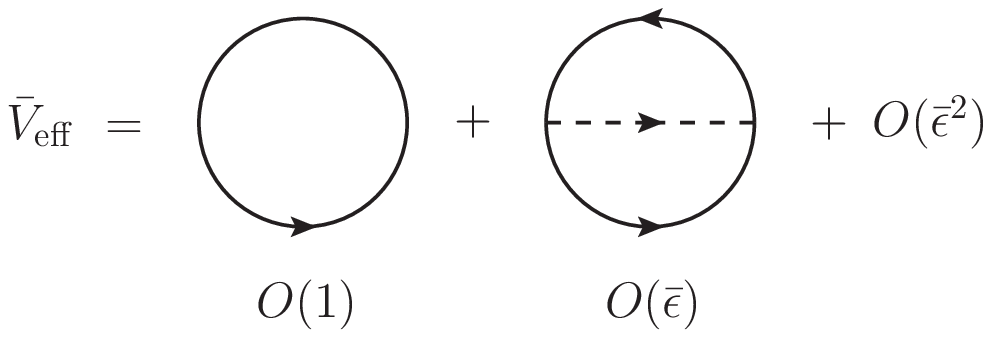}
 \end{center}
 \caption{Vacuum diagrams contributing to the effective potential near
 $d=2$ to leading and next-to-leading orders in $\bar\eps$.
 \label{fig:vacuum_2d}}
\end{figure}

The above result can be further improved by incorporating the expansion
around $d=2$.  Because we can find by solving the gap equation that the
order parameter is exponentially small~\cite{Nishida:2006eu}
\begin{equation}\label{eq:phi_2d}
 \bar\phi_0(\mu) = 2\mu e^{-1/\bar\eps-1+O(\bar\eps)},
\end{equation}
its contribution to the pressure is negligible compared to any powers of
$\bar\eps$.  To leading and next-to-leading orders in $\bar\eps$, the
effective potential receives contributions from two vacuum diagrams
depicted in Fig.~\ref{fig:vacuum_2d}.  Any other diagrams are suppressed
near $d=2$ by further powers of $\bar\eps$.  From the pressure
\begin{equation}
 P = -\bar V_\mathrm{eff}(0)
  = \left[1+\bar\eps+O(\bar\eps^2)\right]
  \frac{2\mu}{\Gamma\!\left(\frac{d}2+2\right)}
  \left(\frac{m\mu}{2\pi}\right)^{d/2},
\end{equation}
we obtain $\xi$ expanded in terms of $\bar\eps$:
\begin{equation}\label{eq:xi_2d}
 \left.\xi\right|_{d\to2} = 1 - \bar\eps + O(\bar\eps^2).
\end{equation}

\begin{figure}[t]
 \begin{center}
  \includegraphics[width=0.49\textwidth]{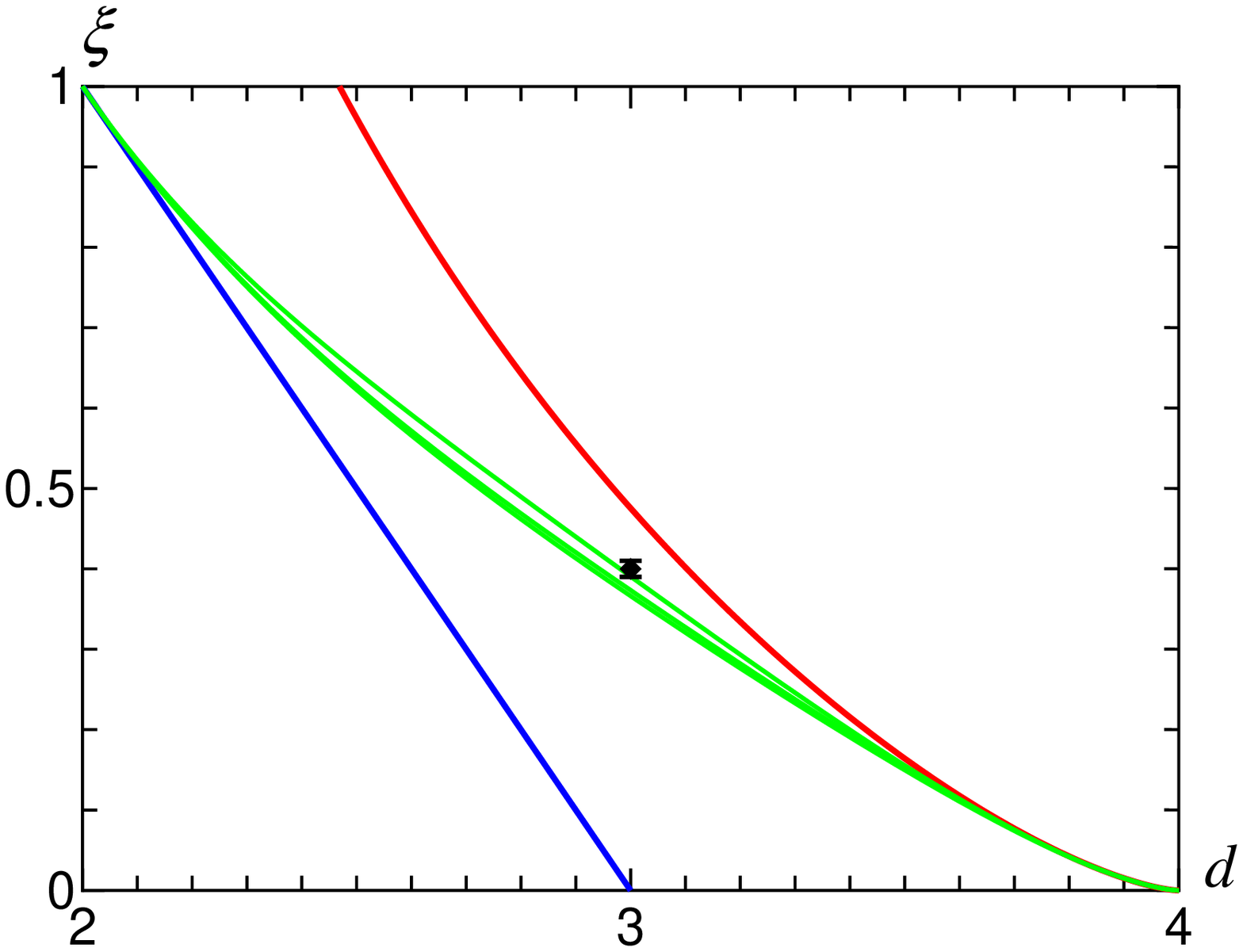}\hfill
  \includegraphics[width=0.49\textwidth]{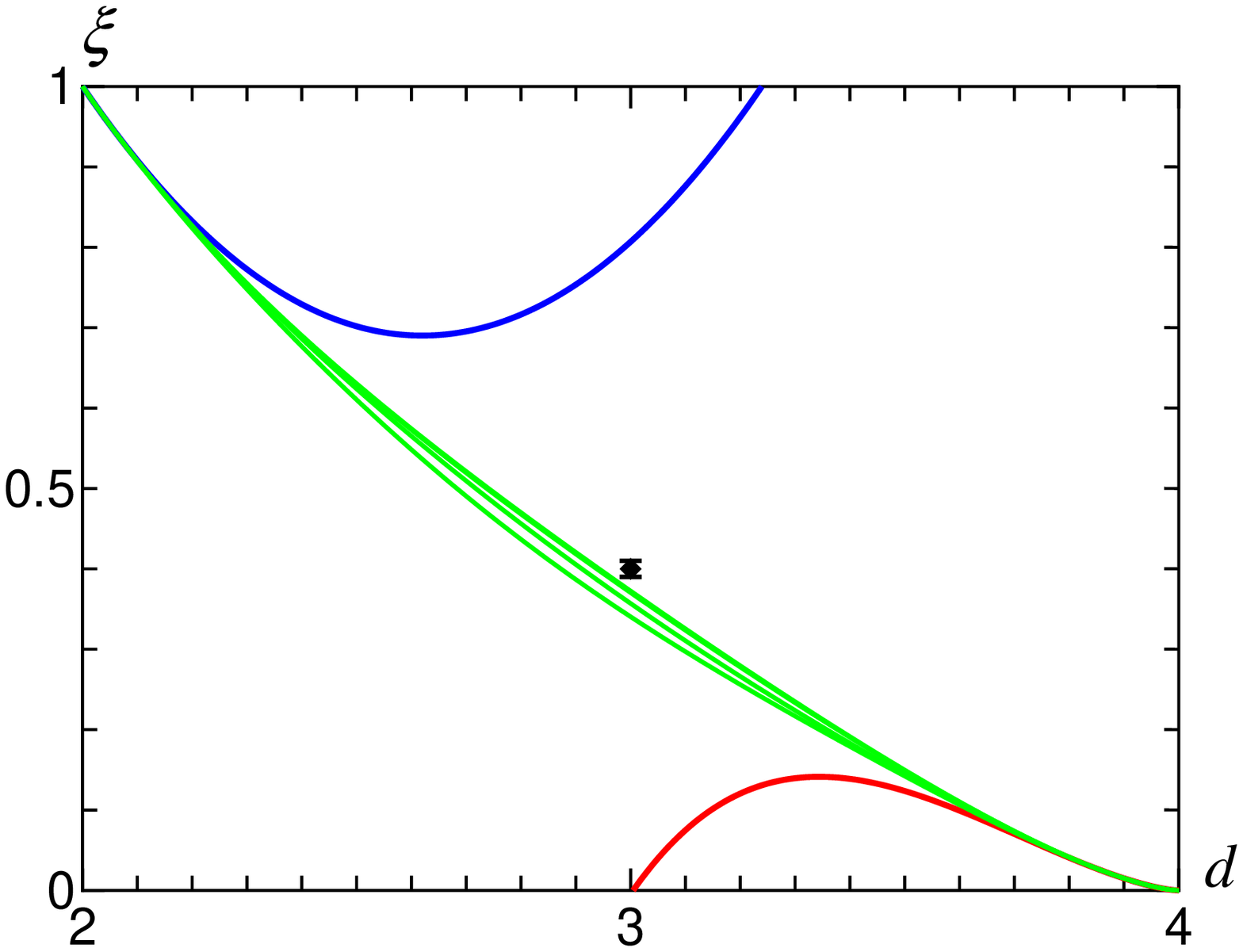}
 \end{center}
 \caption{$\xi$ as a function of the spatial dimension $d$.
 Left panel: The upper (lower) curve is the extrapolation from the NLO
 expansion around $d=4$ in Eq.~(\ref{eq:xi_4d}) [$d=2$ in
 Eq.~(\ref{eq:xi_2d})].  The middle three curves show the Pad\'e
 interpolations of the two NLO expansions.  The symbol at $d=3$
 indicates the result $\xi\approx0.40(1)$ from the Monte Carlo
 simulation~\cite{Zhang}.  Right panel: The same as the left
 panel but including the NNLO corrections in Eqs.~(\ref{eq:nnlo_4d}) and
 (\ref{eq:nnlo_2d}). \label{fig:xi}}
\end{figure}

The value of $\xi$ in $d=3$ can be extracted by interpolating the two
expansions around $d=4$ and $d=2$.  The simplest way to do so is to
employ the Pad\'e approximants.  We write $\xi$ in the form
\begin{equation}\label{eq:Pade_xi}
 \xi = \frac{\eps^{3/2}}2
  \exp\!\left(\frac{\eps\ln\eps}{8-2\eps}\right) F(\eps),
\end{equation}
where $F(\eps)$ is an unknown function and the trivial nonanalytic
dependence on $\eps$ was factored out.\footnote{It has been shown that
$F(\eps)$ has a nonanalytic term $-\frac38\eps^3\ln\eps$ to the
next-to-next-to-next-to-leading order in $\eps$~\cite{Arnold:2006fr}.
Because we are working up to $O(\eps^2)$, we neglect such a nonanalytic
contribution.}  We approximate $F(\eps)$ by ratios of two polynomials
(Pad\'e approximants) and determine their coefficients so that $\xi$ in
Eq.~(\ref{eq:Pade_xi}) has the correct next-to-leading-order (NLO)
expansions both around $d=4$ [Eq.~(\ref{eq:xi_4d})] and $d=2$
[Eq.~(\ref{eq:xi_2d})].  The left panel of Fig.~\ref{fig:xi} shows the
behavior of $\xi$ as a function of $d$.  The middle three curves are the
Pad\'e interpolations of the two NLO expansions.  In $d=3$, these
interpolations give
\begin{equation}\label{eq:nlo_pade}
 \xi \approx 0.391,\ 0.366,\ 0.373,
\end{equation}
which span a small interval $\xi\approx0.377\pm0.014$.  

We can also employ different interpolation schemes, for example, by
applying the Borel transformation to
$F(\eps)=\frac1{\eps}\int_0^\infty\!dt\,e^{-t/\eps}G(t)$ and
approximating the Borel transform $G(t)$ by the Pad\'e
approximants~\cite{Nishida:2006eu}.  These Borel-Pad\'e interpolations
of the two NLO expansions give
\begin{equation}\label{eq:nlo_borel-pade}
 \xi \approx 0.391,\ 0.364,\ 0.378
\end{equation}
in $d=3$, which again span a small interval $\xi\approx0.378\pm0.013$.
Comparing the results in Eqs.~(\ref{eq:nlo_pade}) and
(\ref{eq:nlo_borel-pade}), we find that the interpolated values do not
depend much on the choice of the Pad\'e approximants and also on the
employment of the Borel transformation.

\subsubsection{Next-to-next-to-leading orders \label{sec:NNLO}}
The systematic calculation of $\xi$ has been carried out up to the 
next-to-next-to-leading-order (NNLO) corrections in terms of
$\eps$~\cite{Arnold:2006fr}:
\begin{equation}\label{eq:nnlo_4d}
 \left.\xi\right|_{d\to4} = \frac{\eps^{3/2}}2
  \exp\!\left(\frac{\eps\ln\eps}{8-2\eps}\right)
  \left[1 - 0.04916\,\eps - 0.95961\,\eps^2 + O(\eps^3)\right] 
\end{equation}
and in terms of $\bar\eps$~\cite{Nishida:2008mh,Nishida:2006wk}:
\begin{equation}\label{eq:nnlo_2d}
 \left.\xi\right|_{d\to2}
  = 1 - \bar\eps + 0.80685\,\bar\eps^2 + O(\bar\eps^3).
\end{equation}
The middle four curves in the right panel of Fig.~\ref{fig:xi} are the
Pad\'e interpolations of the two NNLO expansions.  In $d=3$, these
interpolations give
\begin{equation}\label{eq:nnlo_pade}
 \xi \approx 0.340,\ 0.372,\ 0.370,\ 0.357,
\end{equation}
which span an interval $\xi\approx0.360\pm0.020$.\footnote{If we
excluded the interpolation by the simple polynomial ($\xi\approx0.340$)
as was done in Ref.~\cite{Arnold:2006fr}, we would obtain
$0.367\pm0.010$, which is consistent with the Borel-Pad\'e
interpolations without the NNLO correction near
$d=2$~\cite{Arnold:2006fr}.}  In spite of the large NNLO corrections
both near $d=4$ and $d=2$, the interpolated values are roughly
consistent with the previous interpolations of the NLO expansions
(compare the two panels in Fig.~\ref{fig:xi}).  This indicates that the
interpolated results are stable to inclusion of higher-order corrections
and thus the $\eps$ expansion has a certain predictive power even in the
absence of the knowledge on the higher-order corrections.

Finally we note that the limit $\xi|_{d\to4}\to0$ is consistent with the
Nussinovs' picture of the unitary Fermi gas as a noninteracting Bose gas
and $\xi|_{d\to2}\to1$ is consistent with the picture as a
noninteracting Fermi gas in Sect.~\ref{sec:nussinov}.  It would be
interesting to consider how $\xi$ should be continued down to $d\to1$.
The unitary Fermi gas in $d=3$ is analytically continued to
spin-$\frac12$ fermions with a hard-core repulsion in $d=1$, which is
equivalent for the thermodynamic quantities to free identical fermions
with the same total density~\cite{Nikolic:2007zz}.  Therefore it is easy
to find
\begin{equation}
 \left.\xi\right|_{d\to1} \to 4.
\end{equation}
The incorporation of this constraint on the Pad\'e interpolations of the
two NNLO expansions yields $\xi\approx0.365\pm0.010$ in $d=3$.

\subsection{Quasiparticle spectrum \label{sec:spectrum}}
The $\eps$ expansion is also useful to compute other physical
quantities of the unitary Fermi gas.  Here we determine the spectrum of
fermion quasiparticles in a superfluid.  To leading order in $\eps$, the
dispersion relation $\omega(\p)$ is given by the pole of the fermion
propagator in Eq.~(\ref{eq:G_4d});
$\omega(\p)=E_\p=\sqrt{\ep^2+\phi_0^2}$.  It has a minimum at $|\p|=0$
with the energy gap equal to $\Delta=\phi_0=2\mu/\eps$.

\begin{figure}[t]
 \begin{center}
  \includegraphics[width=0.8\textwidth]{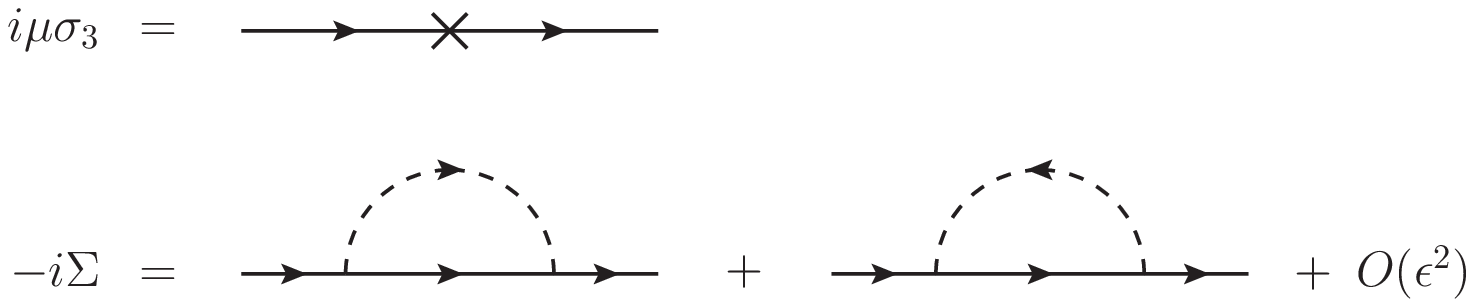}
 \end{center}
 \caption{Fermion self-energy diagrams near $d=4$ to the order
 $O(\eps)$. \label{fig:self-energy_4d}}
\end{figure}

To the next-to-leading order in $\eps$, the fermion propagator receives
corrections from three self-energy diagrams depicted in
Fig.~\ref{fig:self-energy_4d}.  We can find that $\Sigma\sim O(\eps)$ is
diagonal with the elements
\begin{equation}
 \Sigma_{11}(p_0,\p) = -\frac{g^2}2\int\!\frac{d\k}{(2\pi)^d}
  \frac{E_\k-\ek}{E_\k(E_\k+\frac{\varepsilon_{\k-\p}}2-p_0-i0^+)}
\end{equation}
and $\Sigma_{22}(p_0,\p)=-\Sigma_{11}(-p_0,-\p)$.  From the pole of the
fermion propagator,\\
$\det\!\left[G^{-1}(\omega,\p)+\mu\sigma_3-\Sigma(\omega,\p)\right]=0$,
we obtain the dispersion relation
\begin{equation}
 \omega(\p) = \Ep + \frac{\Sigma_{11}+\Sigma_{22}}2
  + \frac{\Sigma_{11}-\Sigma_{22}-2\mu}{2\Ep}\ep + O(\eps^2),
\end{equation}
where $\Sigma_{11}$ and $\Sigma_{22}$ are evaluated at $p_0=E_\p$.

The minimum of $\omega(\p)$ will appear at small momentum
$\ep\sim O(\eps)$.  By expanding $\omega(\p)$ with respect to $\ep$, we
find that the dispersion relation around its minimum has the form
\begin{equation}
 \omega(\p) \simeq \Delta + \frac{(\ep-\varepsilon_0)^2}{2\phi_0}.
\end{equation}
Here $\Delta$ is the energy gap of the fermion quasiparticle given by
\begin{eqnarray}\label{eq:gap}
 \left.\Delta\right|_{d\to4} &=& \phi_0 \nonumber
  \left[1 - \left(8\ln3-12\ln2\right)\eps + O(\eps^2)\right] \\
  &=& \frac{2\mu}\eps
   \left[1 + \left(3C-1-8\ln3+13\ln2\right)\eps + O(\eps^2)\right].
\end{eqnarray}
The minimum of the dispersion relation is located at
$\ep=\varepsilon_0>0$ with
\begin{equation}\label{eq:min}
 \left.\varepsilon_0\right|_{d\to4}
  = \mu + \frac{\eps\phi_0}2 + O(\eps^2) = 2\mu + O(\eps^2),
\end{equation}
where the solution of the gap equation (\ref{eq:phi_4d}) was substituted
to $\phi_0$.

We can see that the next-to-leading-order correction is reasonably
small compared to the leading term at least for
$\Delta/\mu=2/\eps\left(1-0.345\,\eps\right)$.  The naive extrapolation
to the physical dimension $d=3$ gives
\begin{equation}
 \frac{\Delta}{\mu} \to 1.31 \qquad\text{and}\qquad
  \frac{\varepsilon_0}{\mu} \to 2 \qquad (\eps\to1),
\end{equation}
both of which are quite close to the results from the Monte Carlo
simulation $\Delta/\mu\approx1.2$ and
$\varepsilon_0/\mu\approx1.9$~\cite{Carlson:2005kg}.  For comparison,
the mean field approximation yields $\Delta/\mu\approx1.16$ and
$\varepsilon_0/\mu\approx1$.

Frequently $\Delta$ and $\varepsilon_0$ are normalized by the Fermi
energy $\eF$.  By multiplying Eqs.~(\ref{eq:gap}) and (\ref{eq:min}) by
$\mu/\eF=\xi$ obtained previously in Eq.~(\ref{eq:xi_4d}), we find
\begin{equation}
 \frac{\Delta}{\eF} \to 0.604 \qquad\text{and}\qquad
  \frac{\varepsilon_0}{\eF} \to 1 \qquad (\eps\to1).
\end{equation}
For comparison, the Monte Carlo simulation gives $\Delta/\eF\approx0.50$
and $\varepsilon_0/\eF\approx0.8$~\cite{Carlson:2005kg} and the mean
field approximation yields $\Delta/\eF\approx0.686$ and
$\varepsilon_0/\eF\approx0.591$.

\subsection{Critical temperature \label{sec:Tc}}
The formulation of the $\eps$ expansion can be extended to finite
temperature $T\neq0$ by using the imaginary time
prescription~\cite{Nishida:2006rp}.  Here we determine the critical
temperature $\Tc$ for the superfluid phase transition of the unitary
Fermi gas.

The leading contribution to $\Tc/\eF$ in terms of $\eps$ can be obtained
by the following simple argument.  In the limit $d\to4$, the unitary
Fermi gas reduces to a noninteracting fermions and bosons with their
chemical potentials $\mu$ and $2\mu$, respectively [see
Eq.~(\ref{eq:L_4d})].  Because the boson's chemical potential vanishes
at the BEC critical temperature, the density at $T=\Tc$ is given by
\begin{eqnarray}\label{eq:density}
 n &=& 2\int\!\frac{d\p}{(2\pi)^4} \nonumber
  \left[f_\mathrm{F}(\ep)+f_\mathrm{B}(\ep/2)\right] + O(\eps) \\
  &=& \left[\frac{\pi^2}6 + \frac{8\pi^2}6\right]
  \left(\frac{m\Tc}{2\pi}\right)^2 + O(\eps)
  = \frac{3\pi^2}2\left(\frac{m\Tc}{2\pi}\right)^2 + O(\eps).
\end{eqnarray}
Comparing the contributions from Fermi and Bose distributions, we can
see that only 8 of 9 fermion pairs form the composite bosons while 1 of
9 fermion pairs is dissociated.  With the use of
$\eF\equiv\frac{2\pi}{m}\left[\Gamma\!\left(\frac{d}2+1\right)\frac{n}2\right]^{2/d}$
at $d=4$, we obtain
\begin{equation}
 \left.\frac{\Tc}{\eF}\right|_{d\to4} = \sqrt{\frac2{3\pi^2}} + O(\eps).
\end{equation}

\begin{figure}[t]
 \begin{center}
  \includegraphics[width=0.74\textwidth]{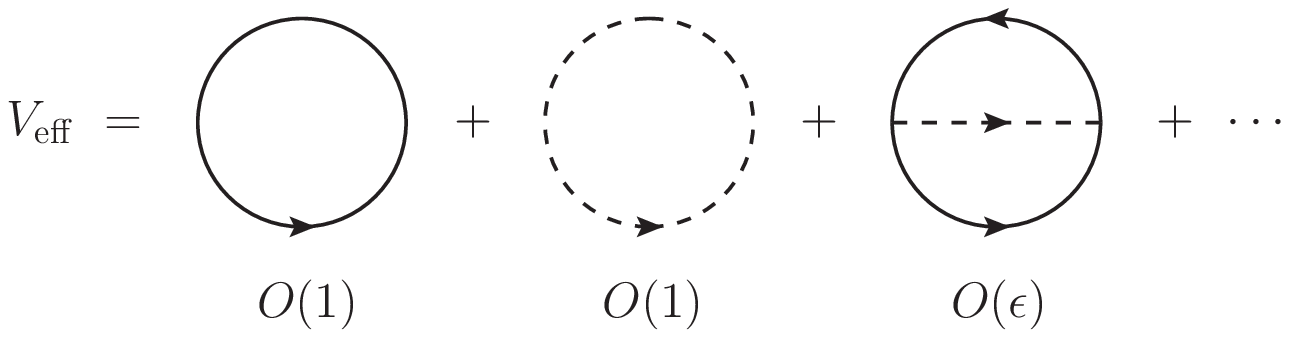}
 \end{center}
 \caption{Three types of diagrams contributing to the effective
 potential at finite temperature near $d=4$ to leading and
 next-to-leading orders in $\eps$.  Each $\mu$ $(2\mu)$ insertion to the
 fermion (boson) line increases the power of $\eps$ by one.  Note that
 at $T\geq\Tc$ where $\phi_0=0$, the $\mu$ insertion to the first
 diagram does not produce the $1/\eps$ singularity.
 \label{fig:finite_T}}
\end{figure}

In order to compute the $O(\eps)$ correction to $\Tc/\eF$, we need to
evaluate the effective potential $\Veff(\phi_0)$ at finite temperature.
The effective potential to leading order in $\eps$ is given by the
fermion one-loop diagrams with and without one $\mu$ insertion and the
boson one-loop diagram.  Near $T\simeq\Tc$, we can expand
$\Veff(\phi_0)$ in $\phi_0/T$ and obtain
\begin{equation}
 \Veff(\phi) - \Veff(0) \simeq
  \left[T\ln2-\frac\mu{\eps}\right]\left(\frac{m\phi_0}{2\pi}\right)^2
  + \frac{\phi_0^2}{16T}\left(\frac{m\phi_0}{2\pi}\right)^2 + \cdots.
\end{equation}
The second order phase transition occurs when the coefficient of the
quadratic term in $\phi_0$ vanishes.  Accordingly, at $T=\Tc$, the
chemical potential is found to be
\begin{equation}\label{eq:mu}
 \mu = \eps\Tc\ln2 + O(\eps^2).
\end{equation}
We then compute the $O(\eps)$ correction to the density
$n=-\d\Veff/\d\mu$ in Eq.~(\ref{eq:density}) at $T=\Tc$ and hence
$\phi_0=0$.  For this purpose, we need to evaluate the fermion and boson
one-loop diagrams with two $\mu$ insertions and the two-loop diagrams
with one $\mu$ insertion to the fermion and boson propagators
(Fig.~\ref{fig:finite_T}).  Performing the loop integrations and
substituting Eq.~(\ref{eq:mu}), we can find
\begin{equation}
 n = \left[\frac{3\pi^2}2-\left\{\frac{3\pi^2\ln2+18\zeta'(2)}4
			   +D-2(\ln2)^2\right\}\eps\right]
 \left(\frac{m\Tc}{2\pi}\right)^{d/2},
\end{equation}
where $D\approx1.92181$ is a numerical constant.  From the definition
$\eF\equiv\frac{2\pi}{m}\left[\Gamma\!\left(\frac{d}2+1\right)\frac{n}2\right]^{2/d}$,
we obtain $\Tc/\eF$ up to the next-leading order in $\eps$:
\begin{equation}\label{eq:Tc_4d}
  \left.\frac{\Tc}{\eF}\right|_{d\to4} = 0.260 - 0.0112\,\eps + O(\eps^2).
\end{equation}

On the other hand, because the unitary Fermi gas near $d=2$ reduces to a
weakly interacting Fermi gas, the critical temperature is provided by
the usual BCS formula $\Tc=\left(e^\gamma/\pi\right)\Delta$, where
$\Delta$ is the energy gap of the fermion quasiparticle at zero
temperature.  Because $\Delta$ in the expansion over $\bar\eps$ is
already obtained by $\bar\phi_0=2\mu e^{-1/\bar\eps-1+O(\bar\eps)}$ in
Eq.~(\ref{eq:phi_2d}), we easily find
\begin{equation}\label{eq:Tc_2d}
 \left.\frac{\Tc}{\eF}\right|_{d\to2}
  = \frac{2e^\gamma}{\pi}e^{-1/\bar\eps-1+O(\bar\eps)},
\end{equation}
where we used $\mu/\eF=\xi=1+O(\eps)$ [Eq.~(\ref{eq:xi_2d})].  The
exponential $e^{-1/\bar\eps}$ is equivalent to the mean field result
while the correction $e^{-1}$ corresponds to the
Gor'kov-Melik-Barkhudarov correction in $d=3$.

\begin{figure}[t]
 \begin{center}
  \includegraphics[width=0.55\textwidth]{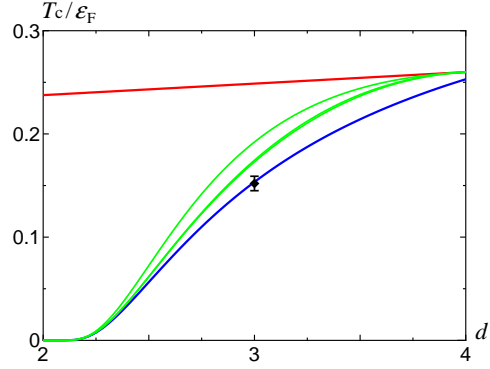}
 \end{center}
 \caption{Critical temperature $\Tc$ as a function of the spatial
 dimension $d$.  The upper (lower) curve is the extrapolation from the
 NLO expansion around $d=4$ in Eq.~(\ref{eq:Tc_4d}) [$d=2$ in
 Eq.~(\ref{eq:Tc_2d})].  The middle three curves show the Pad\'e and
 Borel-Pad\'e interpolations of the two NLO expansions.  The symbol at
 $d=3$ indicates the result $\Tc/\eF=0.152(7)$ from the Monte Carlo
 simulation~\cite{Burovski:2006zz}. \label{fig:Tc}}
\end{figure}

Now the value of $\Tc/\eF$ in $d=3$ can be extracted by interpolating
the two expansions around $d=4$ and $d=2$ just as has been done for
$\xi$ in Sect.~\ref{sec:NLO}.  We write $\Tc/\eF$ in the form
\begin{equation}\label{eq:Pade_Tc}
 \frac{\Tc}{\eF} = 
  \frac{2e^\gamma}{\pi}e^{-1/\bar\eps-1} F(\bar\eps),
\end{equation}
where $F(\bar\eps)$ is an unknown function and the nonanalytic
dependence on $\bar\eps$ was factored out.  We approximate $F(\bar\eps)$
or its Borel transform $G(t)$ by ratios of two polynomials and determine
their coefficients so that $\Tc/\eF$ in Eq.~(\ref{eq:Pade_Tc}) has the
correct NLO expansions both around $d=4$
[Eq.~(\ref{eq:Tc_4d})] and $d=2$ [Eq.~(\ref{eq:Tc_2d})].
Fig.~\ref{fig:Tc} shows the behavior of $\Tc/\eF$ as a function of $d$.
The middle three curves are the Pad\'e and Borel-Pad\'e interpolations
of the two NLO expansions.  In $d=3$, these interpolations give
\begin{equation}
 \frac{\Tc}{\eF} \approx 0.173,\ 0.175,\ 0.192,
\end{equation}
which span an interval $\Tc/\eF\approx0.180\pm0.012$.  Our interpolated
values are not too far from the result of the Monte Carlo simulation
$\Tc/\eF=0.152(7)$~\cite{Burovski:2006zz}.  For comparison, the mean
field approximation yields $\Tc/\eF\approx0.496$.

\subsection{Phase diagram of an imbalanced Fermi gas \label{sec:imbalance}}
The phase diagram of a density- and mass-imbalanced Fermi gas was studied
using the $\eps$ expansion in Refs.~\cite{Nishida:2006eu,Nishida:2006wk}.

\section{Aspects as a nonrelativistic conformal field theory
 \label{sec:NRCFT}}
Because of the absence of scales in the zero-range and infinite scattering
interaction, the theory describing fermions in the unitarity limit is a
nonrelativistic conformal field theory (NRCFT).  In this section, after
deriving the Schr\"odinger algebra and the operator-state correspondence
in NRCFTs, we compute the scaling dimensions of few-body composite
operators exactly or with the help of the $\eps$ expansion, which
provide the energies of a few fermions at unitarity in a harmonic
potential.

\subsection{Schr\"odinger algebra \label{sec:algebra}}
We start with a brief review of the Schr\"odinger
algebra~\cite{Hagen:1972pd,Niederer:1972zz}.  For the latter application,
we allow spin-$\up$ and $\down$ fermions to have different masses
$m_\up$ and $m_\down$.  We define the mass density:
\begin{equation}
 \rho(\x) \equiv \sum_{\sigma=\up,\down}
  m_\sigma\psi_\sigma^\+(\x)\psi_\sigma(\x)
\end{equation}
and the momentum density:
\begin{equation}
 j_i(\x) \equiv -\frac{i}2\sum_{\sigma=\up,\down}
  \left[\psi_\sigma^\+(\x)\d_i\psi_\sigma(\x)
   -\d_i\psi_\sigma^\+(\x)\psi_\sigma(\x)\right],
\end{equation}
where $i=1,\ldots,d$ and the arguments of time are suppressed.  The
Schr\"odinger algebra is formed by the following set of operators:
\begin{eqnarray}
 \text{mass}&:&\quad M \equiv \int\!d\x\,\rho(\x) \\
 \text{momentum}&:&\quad P_i \equiv \int\!d\x\,j_i(\x) \\
 \text{angular momentum}&:&\quad
  J_{ij} \equiv \int\!d\x\left[x_ij_j(\x)-x_jj_i(\x)\right] \\
 \text{Galilean boost}&:&\quad K_i \equiv \int\!d\x\,x_i\rho(\x) \\
 \text{dilatation}&:&\quad D \equiv \int\!d\x\,\x\cdot\bm{j}(\x) \\
 \text{special conformal}&:&\quad C \equiv \int\!d\x\,\frac{\x^2}2\rho(\x)
\end{eqnarray}
and the Hamiltonian:
\begin{eqnarray}
  H &=& \sum_{\sigma=\up,\down}\int\!d\x \nonumber
   \frac{\grad\psi_\sigma^\+(\x)\cdot\grad\psi_\sigma(\x)}{2m_\sigma} \\
  && + \int\!d\x\!\int\!d\y\,\psi_\up^\+(\x)\psi_\down^\+(\y)
  V(|\x-\y|)\psi_\down(\y)\psi_\up(\x).
\end{eqnarray}
$D$ and $C$ are the generators of the scale transformation
($\x\to e^{\lambda}\x,\ t\to e^{2\lambda}t$) and the special conformal
transformation [$\x\to\x/(1+\lambda t),\ t\to t/(1+\lambda t)$],
respectively.  In a scale invariant system such as fermions in the
unitarity limit, these operators form a closed algebra.\footnote{One
potential that realizes the unitarity interaction is
$V(r)=(\pi/2)^2\lim_{r_0\to0}\theta(r_0-r)/(2m_\updown r_0^2)$, where
$m_\updown\equiv m_\up m_\down/(m_\up+m_\down)$ is the reduced mass.}

\begin{table}[t]
 \renewcommand\arraystretch{1.3}
 \caption{Part of the Schr\"odinger algebra.  The values of $[X,\,Y]$
 are shown below. \label{tab:algebra}}
 \begin{center}
  \begin{tabular*}{0.7\textwidth}{@{\extracolsep{\fill}}c|ccccc}
   \hline\hline
   $\ \ X\,\backslash\,Y\ $
   & $P_j$ & $K_j$ & $D$ & $C$ & $H$ \qquad \\
   \hline
   $P_i$ & $0$ & $-i\delta_{ij}M$ & $-iP_i$ & $-iK_i$ & $0$ \qquad \\
   $K_i$ & $i\delta_{ij}M$ & $0$ & $iK_i$ & $0$ & $iP_i$ \qquad \\
   $D$ & $iP_j$ & $-iK_j$ & $0$ & $-2iC$ & $2iH$ \qquad \\
   $C$ & $iK_j$ & $0$ & $2iC$ & $0$ & $iD$ \qquad \\
   $H$ & $0$ & $-iP_j$ & $-2iH$ & $-iD$ & $0$ \qquad \\
   \hline\hline
  \end{tabular*}
 \end{center}
\end{table}

Commutation relations of the above operators are summarized in
Table~\ref{tab:algebra}.  The rest of the algebra is the commutators of
$M$, which commutes with all other operators; $[M,\,\mathrm{any}]=0$.
The commutation relations of $J_{ij}$ with other operators are determined
by their transformation properties under rotations:
\begin{subequations}
 \begin{eqnarray}
  && [J_{ij},\,N] = [J_{ij},\,D] = [J_{ij},\,C] = [J_{ij},\,H] = 0, \\
  && [J_{ij},\,P_k] = i(\delta_{ik} P_j - \delta_{jk} P_i), \qquad
   [J_{ij},\,K_k] = i(\delta_{ik} K_j - \delta_{jk} K_i), \\
  && [J_{ij},\,J_{kl}]= i(\delta_{ik} J_{jl} + \delta_{jl} J_{ik}
   - \delta_{il} J_{jk} - \delta_{jk} J_{il}).
 \end{eqnarray}
\end{subequations}
These commutation relations can be verified by direct calculations,
while only the commutator of $[D,\,H]=2iH$ requires the scale invariance
of the Hamiltonian in which the interaction potential has to satisfy
$V(e^{\lambda}r)=e^{-2\lambda}V(r)$.

\subsection{Operator-state correspondence \label{sec:correspondence}}

\subsubsection{Primary operators \label{sec:primary}}
We then introduce local operators $\O(t,\x)$ as operators that depend
on the position in time and space $(t,\x)$ so that
\begin{equation}
  \O(t,\x) = e^{iHt-iP_ix_i} \O(0) e^{-iHt+i P_ix_i}.
\end{equation}
A local operator $\O$ is said to have a scaling dimension $\Delta_\O$
and a mass $M_\O$ if it satisfies
\begin{equation}
 [D,\,\O(0)] = i\Delta_\O\O(0)
  \qquad\text{and}\qquad
  [M,\,\O(0)] = M_\O\O(0).
\end{equation}

With the use of the commutation relations in Table~\ref{tab:algebra}, we
find that if a given local operator $\O$ has the scaling dimension
$\Delta_\O$, then the scaling dimensions of new local operators
$[P_i,\,\O]$, $[H,\,\O]$, $[K_i,\,\O]$, and $[C,\,\O]$ are given by
$\Delta_\O+1$, $\Delta_\O+2$, $\Delta_\O-1$, and $\Delta_\O-2$,
respectively.  Therefore, by repeatedly taking the commutators with
$K_i$ and $C$, one can keep to lower the scaling dimensions.  However,
this procedure has to terminate because the scaling dimensions of local
operators are bounded from below as we will show in
Sect.~\ref{sec:bound}.  The last operator $\O_\mathrm{pri}$ obtained in
this way must have the property
\begin{equation}
  [K_i,\,\O_\mathrm{pri}(0)] = [C,\,\O_\mathrm{pri}(0)] = 0.
\end{equation}
Such operators that at $t=0$ and $\x=\0$ commute with $K_i$ and $C$ will
be called {\em primary operators\/}.  Below the primary operator
$\O_\mathrm{pri}$ is simply denoted by $\O$.

\begin{figure}[t]
 \begin{center}
  \includegraphics[width=0.67\textwidth]{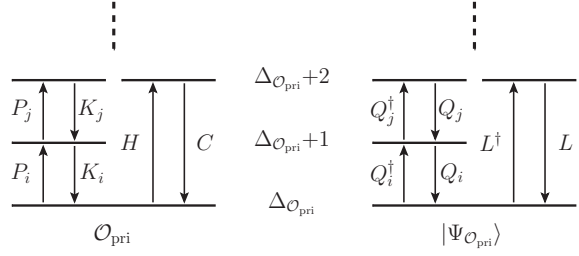}
 \end{center}
 \caption{Correspondence between the spectrum of scaling dimensions of
 local operators in NRCFT (left tower) and the energy spectrum of states
 in a harmonic potential (right tower).  The bottom of each tower
 corresponds to the primary operator $\O_\mathrm{pri}$ (primary state
 $|\Psi_{\O_\mathrm{pri}}\>$). \label{fig:spectrum}}
\end{figure}

Starting with an arbitrary primary operator $\O(t,\x)$, one can build up
a tower of local operators by taking its commutators with $P_i$ and $H$,
in other words, by taking its space and time derivatives (left tower in
Fig.~\ref{fig:spectrum}).  For example, the operators with the scaling
dimension $\Delta_\O+1$ in the tower are $[P_i,\,\O]=i\d_i\O$.  At the
next level with the scaling dimension $\Delta_{\O}+2$, the following
operators are possible; $[P_i,\,[P_j,\,\O]]=-\d_i\d_j\O$ and
$[H,\,\O]=-i\d_t\O$.  Commuting those operators with $K_i$ and $C$, we
can get back the operators into the lower rungs of the tower.  The task
of finding the spectrum of scaling dimensions of all local operators
thus reduces to finding the scaling dimensions of all primary
operators.

It is worthwhile to note that two-point correlation functions of the
primary operators are constrained by the scale and Galilean invariance
up to an overall constant~\cite{Henkel:1993sg}.  For example, the
two-point correlation function of the primary operator $\O$ with its
Hermitian conjugate is given by
\begin{equation}\label{eq:correlation}
 \<T\,\O(t,\x)\O^\+(0)\> \propto \theta(t)\,t^{-\Delta_\O}
  \exp\!\left(i\,M_{\O^\+}\frac{|\x|^2}{2t}\right).
\end{equation}
This formula or its Fourier transform
$\propto\left(-p_0+\frac{\p^2}{2M_{\O^\+}}-i0^+\right)^{\Delta_\O-d/2-1}$
will be useful to read off the scaling dimension $\Delta_\O$.

\subsubsection{Correspondence to states in a harmonic potential}
We now show that each primary operator corresponds to an energy
eigenstate of the system in a harmonic potential.  The Hamiltonian of
the system in a harmonic potential is
\begin{equation}
  H_\omega \equiv H + \omega^2C,
\end{equation}
where $\omega$ is the oscillator frequency.  We consider a primary
operator $\O(t,\x)$ that is composed of annihilation operators in the
quantum field theory so that $\O^\+(t,\x)$ acts nontrivially on the
vacuum: $\O^\+|0\>\neq0$.  We construct the following state using
$\O^\+$ put at $t=0$ and $\x=\0$:
\begin{equation}
 |\Psi_\O\> \equiv e^{-H/\omega}\O^\+(0)|0\>.
\end{equation}
If the mass of $\O^\+$ is $M_{\O^\+}>0$, then $|\Psi_\O\>$ is a mass
eigenstate with the mass eigenvalue $M_{\O^\+}$:
$M|\Psi_\O\>=M_{\O^\+}|\Psi_\O\>$.  Furthermore, with the use of the
commutation relations in Table~\ref{tab:algebra}, it is straightforward
to show that $|\Psi_\O\>$ is actually an energy eigenstate of the
Hamiltonian $H_\omega$ with the energy eigenvalue $E=\Delta_\O\omega$:
\begin{eqnarray}\label{eq:correspondence}
 H_\omega|\Psi_\O\> \nonumber
  &=& \left(H+\omega^2C\right)e^{-H/\omega}\O^\+(0)|0\> \\
  &=& e^{-H/\omega}\left(\omega^2C-i\omega D\right)\O^\+(0)|0\>
  = \omega\Delta_\O|\Psi_\O\>,
\end{eqnarray}
where we used $[C,\,\O^\+(0)]=0$ and $[D,\,\O^\+(0)]=i\Delta_\O$ and the
fact that both $C$ and $D$ annihilate the vacuum.

Starting with the primary state $|\Psi_\O\>$, one can build up a tower
of energy eigenstates of $H_\omega$ by acting the following raising
operators (right tower in Fig.~\ref{fig:spectrum}):
\begin{equation}
 Q_i^\+ = \frac{P_i}{\sqrt{2\omega}}+i\sqrt{\frac{\omega}2}K_i
  \qquad\text{and}\qquad
  L^\+ = \frac{H}{2\omega}-\frac{\omega}2C+\frac{i}2D.
\end{equation}
From the commutation relations $[H_\omega,\,Q_i^\+]=\omega Q_i^\+$ and
$[H_\omega,\,L^\+]=2\omega L^\+$, $Q_i^\+$ or $L^\+$ acting on
$|\Psi_\O\>$ raise its energy eigenvalue by $\omega$ or $2\omega$.  For
example, the states $(Q_i^\+)^n|\Psi_\O\>$ with $n=0,1,2,\ldots$ have
their energy eigenvalues $E=\bigl(\Delta_\O+n\bigr)\omega$ and
correspond to excitations of the center-of-mass motion in the
$i$-direction, while $(L^\+)^n|\Psi_\O\>$ have
$E=\bigl(\Delta_\O+2n\bigr)\omega$ and correspond to excitations of the
breathing mode~\cite{Werner:2006zz}.  The primary state $|\Psi_\O\>$ is
annihilated by the lowering operators $Q_i$ and $L$:
\begin{equation}\label{eq:bottom}
 Q_i|\Psi_\O\>=0 \qquad\text{and}\qquad L|\Psi_\O\>=0.
\end{equation}
Therefore $|\Psi_\O\>$ corresponds to the bottom of each semi-infinite
tower of energy eigenstates, which is the ground state with respect to
the center-of-mass and breathing mode excitations.

The commutation relations of the Hamiltonian and the raising and
lowering operators in the oscillator space are summarized in
Table~\ref{tab:oscillator}.  It is clear from the above arguments and
also from Tables~\ref{tab:algebra} and \ref{tab:oscillator} that the
roles of $(P_i,\,K_i,\,D,\,C,\,H)$ in the free space is now played by
$(Q_i^\+,\,Q_i,\,H_\omega,\,L,\,L^\+)$ in the oscillator space.
Fig.~\ref{fig:spectrum} illustrates the correspondence between the
spectrum of scaling dimensions of local operators in NRCFT and the
energy spectrum of states in a harmonic potential.

\begin{table}[t]
 \renewcommand\arraystretch{1.3}
 \caption{Commutation relations of the Hamiltonian and the raising and
 lowering operators in the oscillator space.  The values of $[X,\,Y]$
 are shown below. \label{tab:oscillator}}
 \begin{center}
  \begin{tabular*}{0.7\textwidth}{@{\extracolsep{\fill}}c|ccccc}
   \hline\hline
   $\ \ X\,\backslash\,Y\ $
   & $Q_j^\+$ & $Q_j$ & $H_\omega/\omega$ & $L$ & $L^\+$ \qquad \\
   \hline
   $Q_i^\+$ & $0$ & $-\delta_{ij}M$ & $-Q_i^\+$ & $-Q_i$ & $0$ \qquad \\
   $Q_i$ & $\delta_{ij}M$ & $0$ & $Q_i$ & $0$ & $Q_i^\+$ \qquad \\
   $H_\omega/\omega$ & $Q_j^\+$ & $-Q_j$ & $0$ & $-2L$ & $2L^\+$ \qquad \\		 
   $L$ & $Q_j$ & $0$ & $2L$ & $0$ & $H_\omega/\omega$ \qquad \\
   $L^\+$ & $0$ & $-Q_j^\+$ & $-2L^\+$ & $-H_\omega/\omega$ & $0$ \qquad \\
   \hline\hline
  \end{tabular*}
 \end{center}
\end{table}

The operator-state correspondence elucidated here allows us to translate
the problem of finding the energy eigenvalues of the system in a
harmonic potential to another problem of finding the scaling dimensions
of primary operators in NRCFT.  In Sect.~\ref{sec:scaling}, we use this
correspondence to compute the energies of fermions at unitarity in a
harmonic potential.  We note that the similar correspondence between
quantities in a harmonic potential and in a free space has been
discussed in the quantum-mechanical language in
Refs.~\cite{Werner:2006zz,Tan:2004}.

\subsubsection{Unitarity bound of scaling dimensions \label{sec:bound}}
As we mentioned in Sect.~\ref{sec:primary}, the scaling dimensions of
local operators are bounded from below.  The lower bound is equal to
$d/2$, which can be seen from the following intuitive physical
argument.  According to the operator-state correspondence, the scaling
dimension of a primary operator is the energy eigenvalue of particles in
a harmonic potential.  The latter can be divided into the center of mass
energy and the energy of the relative motion.  The ground state energy
of the center of mass motion is $\left(d/2\right)\omega$, while the
energy of the relative motion has to be
non-negative~\cite{Werner:2006zz}.  Thus, energy eigenvalues, and hence
operator dimensions, are bounded from below by $d/2$.

More formally, the lower bound can be derived from the requirement of
non-negative norms of states in our theory~\cite{Tachikawa}.  We
consider the primary state $|\Psi_\O\>$ whose mass and energy
eigenvalues are given by $M_{\O^\+}$ and $\Delta_\O\omega$:
\begin{equation}
 M|\Psi_\O\> = M_{\O^\+}|\Psi_\O\> \qquad\text{and}\qquad
  H_\omega|\Psi_\O\> = \Delta_\O\omega|\Psi_\O\>.
\end{equation}
We then construct the following state:
\begin{equation}
 |\Phi\> \equiv \Biggl(L^\+ -
  \sum_{i=1}^d\frac{Q_i^\+Q_i^\+}{2M_{\O^\+}}\Biggr)|\Psi_\O\>,
\end{equation}
and require that it has a non-negative norm $\<\Phi|\Phi\>\geq0$.  With
the use of the commutation relations in Table~\ref{tab:oscillator} and
Eq.~(\ref{eq:bottom}), the norm of $|\Phi\>$ can be computed as
\begin{equation}
 \<\Phi|\Phi\>
  = \left(\Delta_\O-\frac{d}2\right) \<\Psi_\O|\Psi_\O\> \geq 0.
\end{equation}
Therefore we find the lower bound on the scaling dimensions of arbitrary
primary operators:
\begin{equation}\label{eq:unitarity_bound}
 \Delta_\O \geq \frac{d}2.
\end{equation}
The lower bound $d/2$ multiplied by $\omega$ coincides with the ground
state energy of a single particle in a $d$-dimensional harmonic
potential.

When the primary state $|\Psi_\O\>$ saturates the lower bound
$\Delta_\O=d/2$, the vanishing norm of $\<\Phi|\Phi\>=0$ means that the
state itself is identically zero; $|\Phi\>\equiv0$.  Accordingly, the
state created by the corresponding primary operator $\O(t,\x)$ obeys the
free Schr\"odinger equation:
\begin{equation}\label{eq:saturate}
 \left[i\d_t-\frac{\grad^2}{2M_{\O^\+}}\right]\O^\+(t,\x)|0\> = 0.
\end{equation}
In addition to the trivial one-body operator $\psi_\sigma$, we will see
nontrivial examples of primary operators that saturate the lower bound
of the scaling dimensions.

If a theory contains an operator $\O$ with its scaling dimension between
$d/2$ and $(d+2)/2$, then $\O^\dagger\O$ is a relevant deformation:
$\Delta_{\O^\dagger\O}<d+2$.  Therefore, such a theory should contain a
fine tuning in the $\O^\dagger\O$ channel.  We will see this pattern
explicitly below.

\subsubsection{Nonuniversality of $p$-wave resonances}
One consequence of the unitarity bound on operator dimensions is the
impossibility of achieving universality in $p$-wave resonances in three
spatial dimensions.  This issue was raised in connection with the
$\alpha$-$n$ scattering~\cite{Bertulani:2002sz}.  Theoretically, such a
resonance would be described by the following Lagrangian density:
\begin{eqnarray}
 \mathcal{L} &=& \psi_\sigma^\+
  \left(i\d_t + \frac{\grad^2}{2m}\right)\psi_\sigma
  + c_1 \phi_i^\+\left(i\d_t + \frac{\grad^2}{4m}\right)\phi_i
  + c_2 \phi_i^\+\phi_i \\
 &+& \bigl(\psi_\up^\+ \grad_i \psi_\down^\+
  - \grad_i\psi_\up^\+\psi_\down^\+\bigr)\phi_i
  + \phi_i^\+\bigl(\psi_\down\grad_i\psi_\up
  - \grad_i\psi_\down\psi_\up\bigr),
\end{eqnarray}
where $c_1$ and $c_2$ are bare couplings chosen to cancel out the
divergences in the one-loop self-energy of $\phi_i$.  (In contrast to
the $s$-wave resonance case, the loop integral is cubic divergent and
requires two counter terms to regularize, corresponding to two
simultaneous fine tunings of the scattering length and the effective
range.  In dimensional regularization, $c_1=c_2=0$.)  The field $\phi_i$
now has a finite propagator
\begin{equation}\label{eq:phi_i}
 \int\!dtd\x\,e^{ip_0t-i\p\cdot\x}\<T\,\phi_i(t,\x)\phi_i^\+(0)\>
  \propto \left(-p_0+\frac{\p^2}{4m}-i0^+\right)^{-d/2}.
\end{equation}
Such a theory might appear healthy but the scaling dimension of
$\phi_i$, as one can see explicitly by comparing Eq.~(\ref{eq:phi_i})
with the Fourier transform of Eq.~(\ref{eq:correlation}), is
$\Delta_\phi=1$ which is below the unitarity bound of $d/2$ in $d=3$.
(In a free theory $\Delta_\phi=d+1$, but the fine tunings ``reflect''
$\Delta_\phi$ with respect to $(d+2)/2$ so that $\Delta_\phi$ becomes
$1$.)  Thus, $p$-wave resonances cannot be universal in three spatial
dimensions.  The proof given here is more general than that of
Refs.~\cite{Hammer:2009zh}.  Other examples considered in
Refs.~\cite{Hammer:2009zh} can also be analyzed from the light of the
unitarity bound.

\subsection{Scaling dimensions of composite operators \label{sec:scaling}}
All results derived in Sects.~\ref{sec:algebra} and
\ref{sec:correspondence} can be applied to any NRCFTs.  Here we
concentrate on the specific system of spin-$\frac12$ fermions in the
unitarity limit and study various primary operators and their scaling
dimensions.  The simplest primary operator is the one-body operator
$\psi_\sigma(\x)$ whose scaling dimension is trivially given by
\begin{equation}
 \Delta_{\psi_\sigma} = \frac{d}2.
\end{equation}
This value multiplied by $\omega$ indeed matches the ground state energy
of one fermion in a $d$-dimensional harmonic potential.

\subsubsection{Two-body operator}
The first nontrivial primary operator is the two-body composite
operator: $\phi(\x)\equiv c_0\psi_\down\psi_\up(\x)$, which also appears
as an auxiliary field of the Hubbard-Stratonovich transformation
(\ref{eq:Hubbard-Stratonovich1}).  The presence of the prefactor
$c_0\sim\Lambda^{2-d}$ guarantees that matrix elements of $\phi(\x)$
between two states in the Hilbert space are finite.  Accordingly its
scaling dimension becomes
\begin{equation}\label{eq:D_phi}
 \Delta_{\phi} = \Delta_{\psi_\up}+\Delta_{\psi_\down}+(2-d) = 2.
\end{equation}
This result can be confirmed by computing the two-point correlation
function of $\phi$ and comparing it with the Fourier transform of
Eq.~(\ref{eq:correlation}):
\begin{equation}\label{eq:phi_propagator}
 \int\!dtd\x\,e^{ip_0t-i\p\cdot\x}\<T\,\phi(t,\x)\phi^\+(0)\>
  = \frac{\left(\frac{2\pi}{m_\updown}\right)^{d/2}}
  {\Gamma\!\left(1-\frac{d}2\right)
  \left[-p_0+\frac{\p^2}{2(m_\up+m_\down)}-i0^+\right]^{d/2-1}}.
\end{equation}

According to the operator-state correspondence, the ground state energy
of two fermions at unitarity in a harmonic potential is exactly
$2\omega$ for an arbitrary spatial dimension $d$.  This result is
consistent with our intuitive pictures of spin-$\up$ and $\down$
fermions in the unitarity limit as a single point-like composite boson
in $d=4$, two noninteracting fermions in $d=2$, and two identical
fermions in $d=1$ (see discussions in Sects.~\ref{sec:nussinov} and
\ref{sec:NNLO}).  Note that $\phi(\x)$ in $d=4$ saturates the lower
bound (\ref{eq:unitarity_bound}) and thus obeys the free Schr\"odinger
equation (\ref{eq:saturate}) with mass $M_{\phi^\+}=m_\up+m_\down$.  The
same result in $d=3$ has been obtained by directly solving the two-body
Schr\"odinger equation with a harmonic potential~\cite{Busch:1998},
which is consistent with the experimental
measurement~\cite{Stoferle:2006}.

\subsubsection{Three-body operators \label{sec:3-body}}
We then consider three-body composite operators.  The formula to compute
their scaling dimensions for arbitrary mass ratio $m_\up/m_\down$,
angular momentum $l$, and spatial dimension $d$ is derived in Appendix.
Here we discuss its physical consequences in $d=3$.  Three-body
operators composed of two spin-$\up$ and one spin-$\down$ fermions with
orbital angular momentum $l=0$ and $l=1$ are
\begin{equation}
 \O_{\up\up\down}^{(l=0)}(\x) \equiv Z_0^{-1}\phi\psi_\up(\x)
\end{equation}
and
\begin{equation}\label{eq:l=1}
 \O_{\up\up\down}^{(l=1)}(\x) \equiv Z_1^{-1}
  \left[\left(m_\up+m_\down\right)\phi\d_i\psi_\up
   -m_\up(\d_i\phi)\psi_\up\right](\x),
\end{equation}
where $i=1,2,3$ and $Z_l\sim\Lambda^{-\gamma_l}$ is the renormalization
factor.  The mass factors in Eq.~(\ref{eq:l=1}) are necessary so that
the operator is primary; $[K_i,\O_{\up\up\down}^{(l=1)}(0)]=0$.

\begin{figure}[t]
 \includegraphics[width=0.48\textwidth,clip]{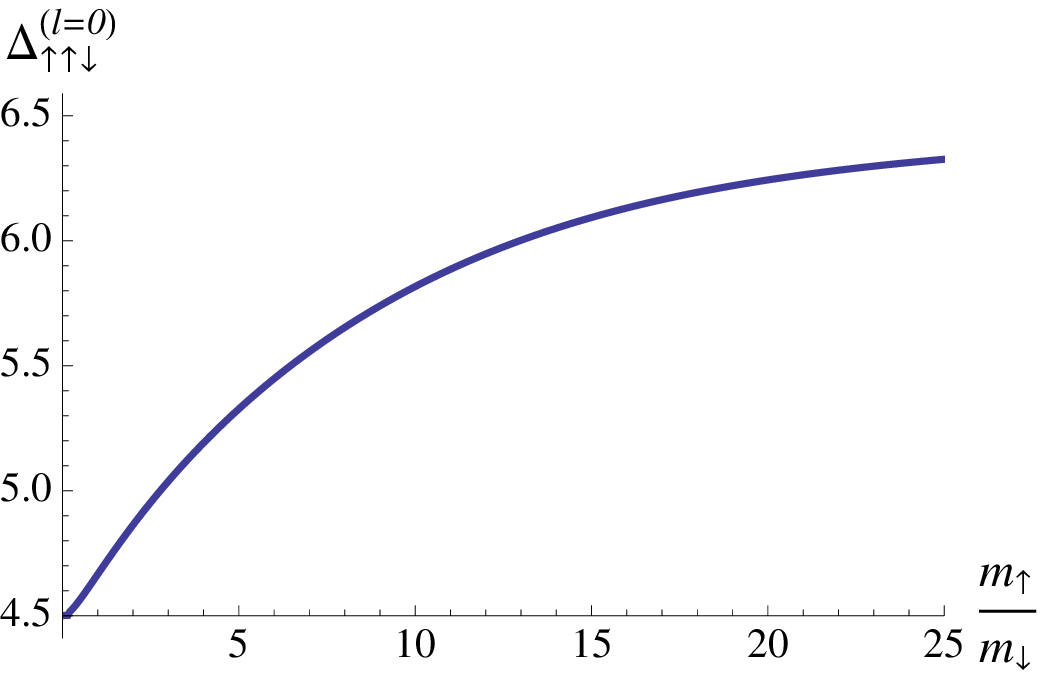}\hfill
 \includegraphics[width=0.48\textwidth,clip]{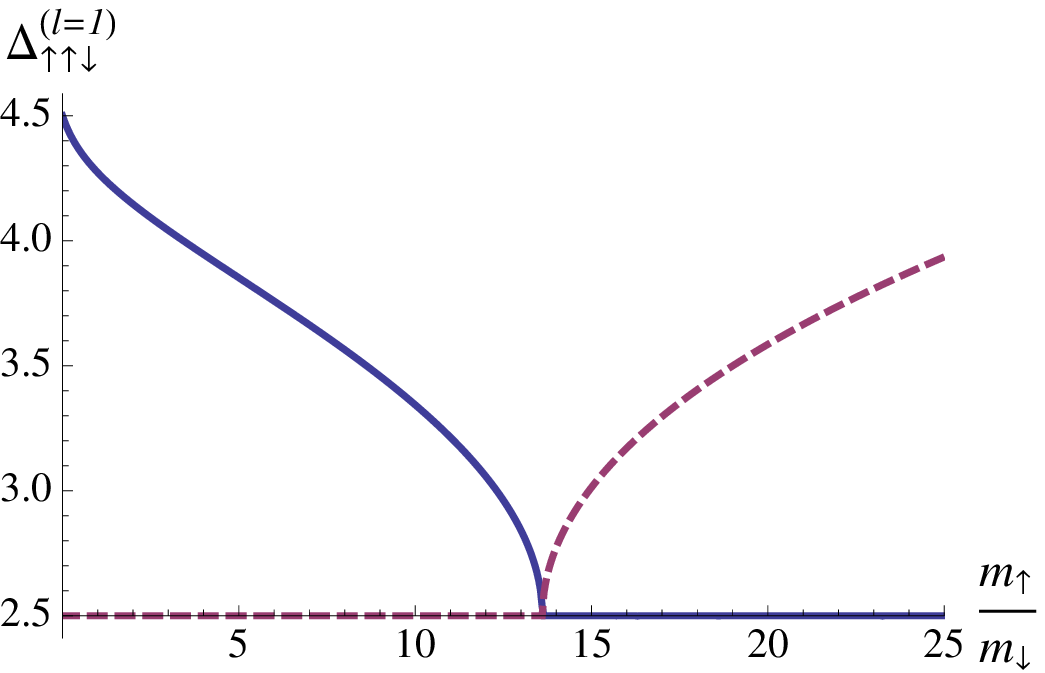}
 \caption{Scaling dimensions of three-body composite operators
 $\O_{\up\up\down}^{(l)}$ with angular momentum $l=0$ (left panel) and
 $l=1$ (right panel) as functions of the mass ratio $m_\up/m_\down$.  In
 the right panel, the real part (solid curve) and the imaginary part
 shifted by ${+}2.5$ (dashed curve) are plotted. \label{fig:mass_ratio}}
\end{figure}

Their scaling dimensions $\Delta_{\up\up\down}^{(l)}=\frac72+l+\gamma_l$
obtained by solving Eq.~(\ref{eq:anomalous}) are plotted in
Fig.~\ref{fig:mass_ratio} as functions of the mass ratio
$m_\up/m_\down$.  For $l=0$ (left panel), $\Delta_{\up\up\down}^{(l)}$
increases as $m_\up/m_\down$ is increased indicating the stronger
effective repulsion in the $s$-wave channel.  
On the other hand, $\Delta_{\up\up\down}^{(l)}$ for $l=1$ (right panel)
decreases with increasing $m_\up/m_\down$ and eventually, when the mass
ratio exceeds the critical value $m_\up/m_\down>13.607$, it becomes
complex as $\Delta_{\up\up\down}^{(l=1)}=\frac52\pm i\,\mathrm{Im}\,
(\gamma_1)$.\footnote{This situation illustrates a general feature:
onsets of the Efimov effect occur when the ground state energy of the
corresponding few-body system in a harmonic potential is equal to
$\frac{d+2}2\omega$.}  In this case, the Fourier transform of
Eq.~(\ref{eq:correlation}) implies that the two-point correlation
function of $\O_{\up\up\down}^{(l=1)}$ behaves as
\begin{equation}
 \propto \sin\!\left[\mathrm{Im}(\gamma_1)
		\ln\left(\frac{\p^2-(4m_\up+2m_\down)p_0-i0^+}
		    {\Lambda^2}\right)+\theta\right].
\end{equation}
Now the full scale invariance of the original NRCFT is broken down to a
discrete scaling symmetry,
\begin{equation}
 \p\to e^{n\pi/\mathrm{Im}(\gamma_1)}\,\p
  \qquad\text{and}\qquad
  p_0\to e^{2n\pi/\mathrm{Im}(\gamma_1)}\,p_0,
\end{equation}
with $n$ being an integer.  This is a characteristic of the
renormalization-group limit cycle and related to the existence of an
infinite set of three-body bound states in the $p$-wave channel.  Their
energy eigenvalues form a geometric spectrum
$E_{n+1}/E_n=e^{-2\pi/|\mathrm{Im}(\gamma_1)|}$, which is known as the
Efimov effect~\cite{Efimov:1972}.  Because the system develops deep
three-body bound states, the corresponding many-body system cannot be
stable toward collapse.

We note that, in the range of the mass ratio
$8.6186<m_\up/m_\down<13.607$, the scaling dimension of
$\O_{\up\up\down}^{(l=1)}$ satisfies
$\frac52<\Delta_{\up\up\down}^{(l=1)}<\frac72$. Accordingly the
following three-body interaction term
\begin{equation}
 \mathcal{L}_\text{3-body} = c_1\,
  \O_{\up\up\down}^{(1)\+}(t,\x)\O_{\up\up\down}^{(1)}(t,\x)
\end{equation}
becomes renormalizable because now the coupling has the dimension
$-2<[c_1]<0$.  The Lagrangian density (\ref{eq:Hubbard-Stratonovich1})
with $\mathcal{L}_\text{3-body}$ added defines a new renormalizable
theory.  In particular, when the coupling $c_1$ is tuned to its
nontrivial fixed point ($c_1\neq0$) that describes a three-body
resonance in the $p$-wave channel, the resulting theory provides a novel
NRCFT.  The corresponding system is that of spin-$\frac12$ fermions with
both two-body ($\up\down$) and three-body ($\up\up\down$)
resonances and its many-body physics was studied in
Ref.~\cite{Nishida:2007mr}.

\subsection[Application of the $\epsilon$ expansion]{Application of the $\bfepsilon$ expansion \label{sec:application}}
It would be difficult to determine exact scaling dimensions of composite
operators with more than three fermions in $d=3$.  However, it is
possible to estimate them with the help of the $\eps$ expansions around
$d=4$ and $d=2$.  The formulations developed in
Sect.~\ref{sec:formulation} can be used for the few-body problems as
well just by setting $\mu=\phi_0=0$.  Here we concentrate on the equal
mass case $m_\up=m_\down$.

\subsubsection{Scaling dimensions near four spatial dimensions}
\begin{figure}[t]
 \begin{center}
  \includegraphics[width=0.9\textwidth,clip]{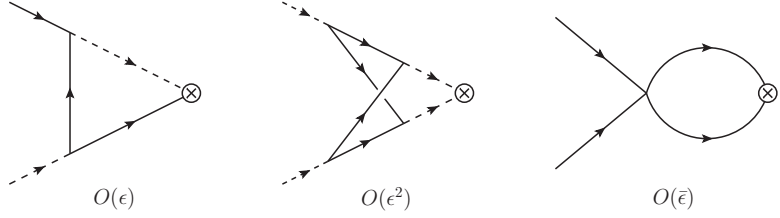}
 \end{center} 
 \caption{Leading-order Feynman diagrams to renormalize the composite
 operators $\phi\psi_\up$ (left) and $\phi\phi$ (middle) near $d=4$ and
 $\psi_\up\psi_\down$ (right) near $d=2$.  These operators are inserted
 at $\otimes$. \label{fig:renormalization}}
\end{figure}

The scaling dimensions of composite operators can be determined by
studying their renormalizations.  We start with the simplest three-body
operator $\phi\psi_\up$ near $d=4$, which has zero orbital angular
momentum $l=0$.  The leading-order diagram that renormalizes
$\phi\psi_\up$ is $O(\eps)$ and depicted in
Fig.~\ref{fig:renormalization} (left).  Performing the loop integration,
we find that this diagram is logarithmically divergent at $d=4$ and thus
the renormalized operator differs from the bare one by the
renormalization factor;
$(\phi\psi_\up)_\mathrm{ren}=Z_{\phi\psi_\up}^{-1}\phi\psi_\up$, where
$Z_{\phi\psi_\up}=1-\frac43\eps\,\ln\Lambda$.  From the anomalous
dimension 
$\gamma_{\phi\psi_\up}=-\ln Z_{\phi\psi_\up}/\ln\Lambda=\frac43\eps$,
we obtain the scaling dimension of the renormalized operator up to the
next-to-leading order in $\eps$:
\begin{equation}\label{eq:3-fermions_s-wave_4d}
 \Delta_{\phi\psi_\up}
  = \Delta_\phi + \Delta_{\psi_\up} + \gamma_{\phi\psi_\up}
  = 4 + \frac56\eps + O(\eps^2).
\end{equation}
According to the operator-state correspondence, there is a three-fermion
state in a harmonic potential with $l=0$ and energy equal to
$E_3^{(0)}=\Delta_{\phi\psi_\up}\omega$, which continues to the first
excited state in $d=3$.  Even within the leading correction in $\eps$,
the naive extrapolation to $\eps\to1$ yields $E_3^{(0)}\to4.83\,\omega$,
which is already close to the true first excited state energy
$4.66622\,\omega$ in $d=3$~\cite{Tan:2004}.

The ground state of three fermions in a harmonic potential has $l=1$ in
$d=3$.  The corresponding primary operator near $d=4$ is
$2\phi\grad\psi_\up-(\grad\phi)\psi_\up$, which is renormalized by the
same diagram in Fig.~\ref{fig:renormalization} (left).  From its
anomalous dimension $\gamma=-\ln Z/\ln\Lambda=-\frac13\eps$, we obtain
the scaling dimension of the renormalized operator up to the
next-to-leading order in $\eps$:
\begin{equation}\label{eq:3-fermions_p-wave_4d}
 \Delta_{2\phi\grad\psi_\up-(\grad\phi)\psi_\up}
  = \Delta_\phi + \Delta_{\psi_\up} + 1 + \gamma
  = 5 - \frac56\eps + O(\eps^2).
\end{equation}
The operator-state correspondence tells us that the three-fermion state
in a harmonic potential with $l=1$ has the energy
$E_3^{(1)}=\Delta_{2\phi\grad\psi_\up-(\grad\phi)\psi_\up}\omega$.  The
naive extrapolation to $\eps\to1$ yields
$E_3^{(1)}\to4.17\,\omega$, which is already close to the true ground
state energy $4.27272\,\omega$ in $d=3$~\cite{Tan:2004}.

We now turn to the lowest four-body operator $\phi\phi$ with zero
orbital angular momentum $l=0$.  The first nontrivial correction to its
scaling dimension is $O(\eps^2)$ given by the diagram depicted in
Fig.~\ref{fig:renormalization} (middle).  The two-loop integral can be
done analytically and we obtain the renormalization factor;
$Z_{\phi\phi}=1-8\eps^2\ln\frac{27}{16}\ln\Lambda$.  Therefore we find
the scaling dimension of the renormalized operator
$(\phi\phi)_\mathrm{ren}=Z_{\phi\phi}^{-1}\phi\phi$ to be
\begin{equation}\label{eq:4-fermion_4d}
 \Delta_{\phi\phi} = 2\Delta_{\phi} + \gamma_{\phi\phi}
  = 4 + 8\eps^2\ln\frac{27}{16} + O(\eps^3).
\end{equation}
According to the operator-state correspondence, the ground state of four
fermions in a harmonic potential has the energy
$E_4^{(0)}=\Delta_{\phi\phi}\omega$.  Because the $O(\eps^2)$ correction
turns out to be large, we shall not directly extrapolate
Eq.~(\ref{eq:4-fermion_4d}) to $\eps\to1$ but will combine it with an
expansion near $d=2$.

The above results can be easily extended to a general number of
fermions by evaluating the diagrams in Fig.~\ref{fig:renormalization}
with more boson lines attached.  The primary operators having $N$
fermion number and orbital angular momentum $l$ are summarized in
Table~\ref{tab:operator_4d} with their scaling dimensions computed in
the $\eps$ expansion.  The energy of the corresponding state in a
harmonic potential is simply given by $E_N^{(l)}=\Delta_\O\omega$.
The leading-order results [$E_N^{(0)}=N\omega$ for even $N$ and
$E_N^{(0)}=(N+1)\omega$ and $E_N^{(1)}=(N+2)\omega$ for odd $N$] can be
intuitively understood by recalling that fermion pairs at unitarity in
$d=4$ form point-like bosons and they do not interact with each other or
with extra fermions.  Therefore the ground state of $N=2n$ fermions
consists of $n$ free composite bosons occupying the same lowest energy
state in a harmonic potential with the energy $2\omega$ in $d=4$.  When
$N=2n+1$, the ground state has $l=0$ and consists of $n$ composite
bosons and one extra fermion occupying the same lowest energy state
again.  In order to create an $l=1$ state, one of the $n+1$ particles
has to be excited to the first excited state, which costs the additional
energy $1\omega$.  The leading correction to the energy, which is
represented by the anomalous dimension in NRCFT, originates from the
weak boson-fermion $[O(\eps)]$ or boson-boson $[O(\eps^2)]$
interaction.  Finally we note that in $d=4$, we can observe the odd-even
staggering in the ground state energy;
$E_N^{(0)}-\bigl(E_{N-1}^{(0)}+E_{N+1}^{(0)}\bigr)/2=1\omega$ for odd
$N$.

\begin{table}[t]
 \renewcommand\arraystretch{1.5}
 \caption{$N$-body composite operators with angular momentum $l$ near
 $d=4$ and their scaling dimensions in the $\eps$ expansion.
 \label{tab:operator_4d}}
 \begin{center}
  \begin{tabular*}{0.9\textwidth}{@{\extracolsep{\fill}}l|l|l}
   \hline\hline
   $\ \ N\ \ \,(l)$ & $\O_N^{(l)}$ & $\Delta_\O$ \\
   \hline
   $\ \ 2n\ \,(l=0)$ & $\phi^n$
       & $N+N(N-2)\eps^2\ln\frac{27}{16}+O(\eps^3)\quad$ \\
   $\ \ 2n+1\ \,(l=0)$ & $\phi^n\phi_\up$
       & $N+1+\frac{4N-7}6\eps+O(\eps^2)$ \\
   $\ \ 2n+1\ \,(l=1)\ \ $
   & $2\phi^n\grad\psi_\up-\phi^{n-1}(\grad\phi)\psi_\up\quad$
       & $N+2+\frac{2N-21}{18}\eps+O(\eps^2)$ \\
   \hline\hline
  \end{tabular*}
 \end{center}
\end{table}

\subsubsection{Scaling dimensions near two spatial dimensions}
Similarly, we can determine the scaling dimensions of composite
operators near $d=2$ by studying their renormalizations.  Here we
consider the three-body operators $\psi_\up\psi_\down\d_t\psi_\up$ and
$\psi_\up\psi_\down\grad\psi_\up$, which are primary and have the
orbital angular momentum $l=0$ and $l=1$, respectively.  The
leading-order diagrams that renormalize them are $O(\bar\eps)$ and
depicted in Fig.~\ref{fig:renormalization} (right) with one more fermion
line attached.  Performing the loop integrations, we find that these
diagrams are logarithmically divergent at $d=2$ and thus the
renormalized operators differ from the bare ones by the renormalization
factors;
$Z_{\psi_\up\psi_\down\d_t\psi_\up}=1+\frac32\bar\eps\ln\Lambda$ and
$Z_{\psi_\up\psi_\down\grad\psi_\up}=1+\frac32\bar\eps\ln\Lambda$.  From
the anomalous dimensions 
$\gamma_\O=-\ln Z_\O/\ln\Lambda=-\frac32\bar\eps$,  we obtain the
scaling dimensions of the renormalized operators up to the
next-to-leading order in $\bar\eps$:
\begin{equation}\label{eq:3-fermions_s-wave_2d}
 \Delta_{\psi_\up\psi_\down\d_t\psi_\up}
  = 3\Delta_{\psi_\sigma} + 2 + \gamma_{\psi_\up\psi_\down\d_t\psi_\up}
  = 5 + O(\bar\eps^2)
\end{equation}
and
\begin{equation}\label{eq:3-fermions_p-wave_2d}
 \Delta_{\psi_\up\psi_\down\grad\psi_\up}
  = 3\Delta_{\psi_\sigma} + 1 + \gamma_{\psi_\up\psi_\down\grad\psi_\up}
  = 4 + O(\bar\eps^2).
\end{equation}
The operator-state correspondence tells us that the three-fermion states
in a harmonic potential with $l=0$ and $l=1$ have the energies
$E_3^{(0)}=\Delta_{\psi_\up\psi_\down\d_t\psi_\up}\omega$ and
$E_3^{(1)}=\Delta_{\psi_\up\psi_\down\grad\psi_\up}\omega$,
respectively.

\begin{table}[t]
 \renewcommand\arraystretch{1.4}
 \caption{$N$-body composite operators with angular momentum $l$ near
 $d=2$ and their scaling dimensions in the $\bar\eps$ expansion.  Known
 values for the energies of $N$ fermions in a harmonic potential in
 $d=3$ are also shown in units of $\hbar\omega$.
 \label{tab:operator_2d}}
 \begin{center}
  \begin{tabular*}{0.95\textwidth}{@{\extracolsep{\fill}}l|l|l|l}
   \hline\hline
   $\ \ N\ (l)$ & $\O_N^{(l)}$ & $\Delta_\O$
   & $E/\hbar\omega\ $ in\, $d=3\ \ $ \\
   \hline
   $\ \ 2\ \ (l=0)\ \ $ & $\psi_\up\psi_\down$
       & $2$ & $2^{~\scriptsize\cite{Busch:1998}}$ \\
   $\ \ 3\ \ (l=0)\ \ $ & $\psi_\up\psi_\down(\d_t\psi_\up)$
       & $5+O(\bar\eps^2)$ & $4.66622^{~\scriptsize\cite{Tan:2004}}$ \\
   $\ \ 3\ \ (l=1)\ \ $ & $\psi_\up\psi_\down(\grad\psi_\up)$
       & $4+O(\bar\eps^2)$ & $4.27272^{~\scriptsize\cite{Tan:2004}}$ \\
   $\ \ 4\ \ (l=0)\ \ $
   & $\psi_\up\psi_\down(\grad\psi_\up{\cdot}\grad\psi_\down)$
       & $6-\bar\eps+(\bar\eps^2)$
	   & $\approx5.028^{~\scriptsize\cite{Blume:2007}}$ \\
   $\ \ 5\ \ (l=0)\ \ $ & $(*)$
       & $9-\frac{11\pm\sqrt{105}}{16}\bar\eps+O(\bar\eps^2)\ \ $
	   & $\approx8.03^{~\scriptsize\cite{Blume:2007}}$ \\	      
   $\ \ 5\ \ (l=1)\ \ $
   & $\psi_\up\psi_\down(\grad\psi_\up{\cdot}\grad\psi_\down)\grad\psi_\up\ \ $
       & $8-\bar\eps+O(\bar\eps^2)$
	   & $\approx7.53^{~\scriptsize\cite{Blume:2007}}$ \\
   $\ \ 6\ \ (l=0)\ \ $
   & $\psi_\up\psi_\down(\grad\psi_\up{\cdot}\grad\psi_\down)^2$
       & $10-2\bar\eps+(\bar\eps^2)$
	   & $\approx8.48^{~\scriptsize\cite{Blume:2007}}$ \\
   \hline\hline
  \end{tabular*}
 \end{center}
 $(*)=a\,\psi_\up\psi_\down(\grad\psi_\up{\cdot}\grad\psi_\down)\d^2\psi_\up+b\,\psi_\up\d_i\psi_\down(\grad\psi_\up{\cdot}\grad\psi_\down)\d_i\psi_\up+c\,\psi_\up\psi_\down((\d_i\grad\psi_\up){\cdot}\grad\psi_\down)\d_i\psi_\up-d\,\psi_\up\psi_\down(\grad\psi_\up{\cdot}\grad\psi_\down)i\d_t\psi_\up$
 with
 $(a,b,c,d)=(\pm19\sqrt{3}-5\sqrt{35},\mp16\sqrt{3},-6\sqrt{35}\mp6\sqrt{3},16\sqrt{35})$.
\end{table}

We can develop the same analysis for composite operators with more than
three fermions by evaluating the diagram in
Fig.~\ref{fig:renormalization} (right) with more fermion lines attached.
The primary operators having $N\leq6$ fermion number and orbital angular
momentum $l$ are summarized in Table~\ref{tab:operator_2d} with their
scaling dimensions computed in the $\bar\eps$ expansion.  Note that
composite operators having the same classical dimension can mix under
the renormalization and thus the primary operator with the well-defined
scaling dimension may have a complicated form such as for $N=5$ and
$l=0$.  The leading order results for the corresponding energies
$E_N^{(l)}=\Delta_\O\omega$ in a harmonic potential can be easily
understood by recalling that fermions at unitarity become noninteracting
in $d=2$.  Therefore the energy eigenvalue of each $N$-fermion state is
just a sum of single particle energies in a harmonic potential in $d=2$,
and obviously, the ground state energy shows the shell structure.  The
$O(\bar\eps)$ correction to the energy, which is represented by the
anomalous dimension in NRCFT, originates from the weak fermion-fermion
interaction.  We can see in Table~\ref{tab:operator_2d} the rough
agreement of the naive extrapolations of $\Delta_\O$ to $\bar\eps\to1$
with the known values in $d=3$.

\subsubsection[Interpolations of $\epsilon$ expansions]{Interpolations of $\bfepsilon$ expansions}
We now extract the energy of $N$ fermions in a harmonic potential in
$d=3$ by interpolating the two expansions around $d=4$ and $d=2$ just as
has been done for $\xi$ in Sect.~\ref{sec:zero_T} and $\Tc/\eF$ in
Sect.~\ref{sec:Tc}.  We approximate $E_N^{(l)}/\omega$ by ratios of two
polynomials (Pad\'e approximants) and determine their unknown
coefficients so that the correct expansions both around $d=4$
(Table~\ref{tab:operator_4d}) and $d=2$ (Table~\ref{tab:operator_2d})
are reproduced.

\begin{figure}[t]
 \includegraphics[width=0.5\textwidth,clip]{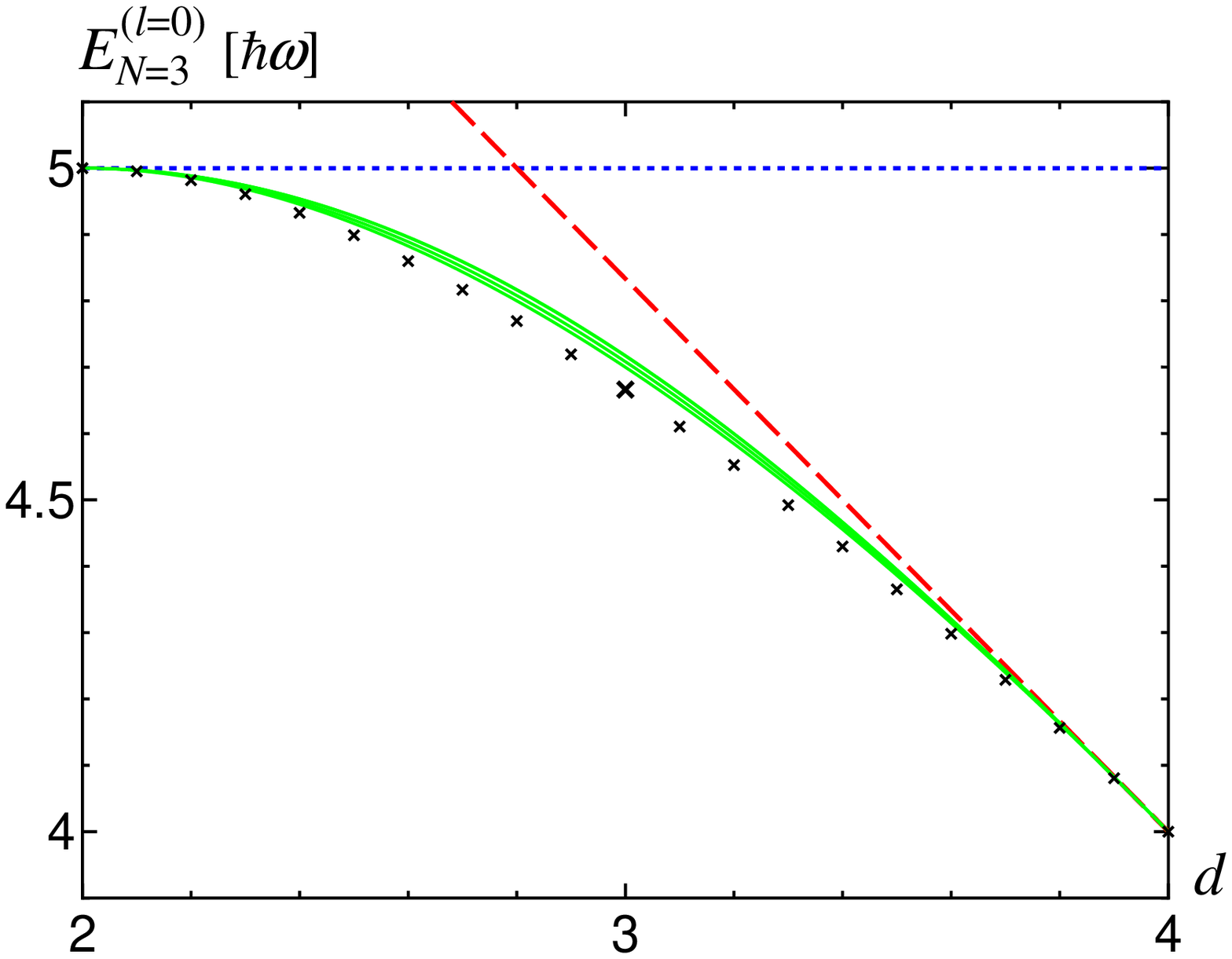}\hfill
 \includegraphics[width=0.5\textwidth,clip]{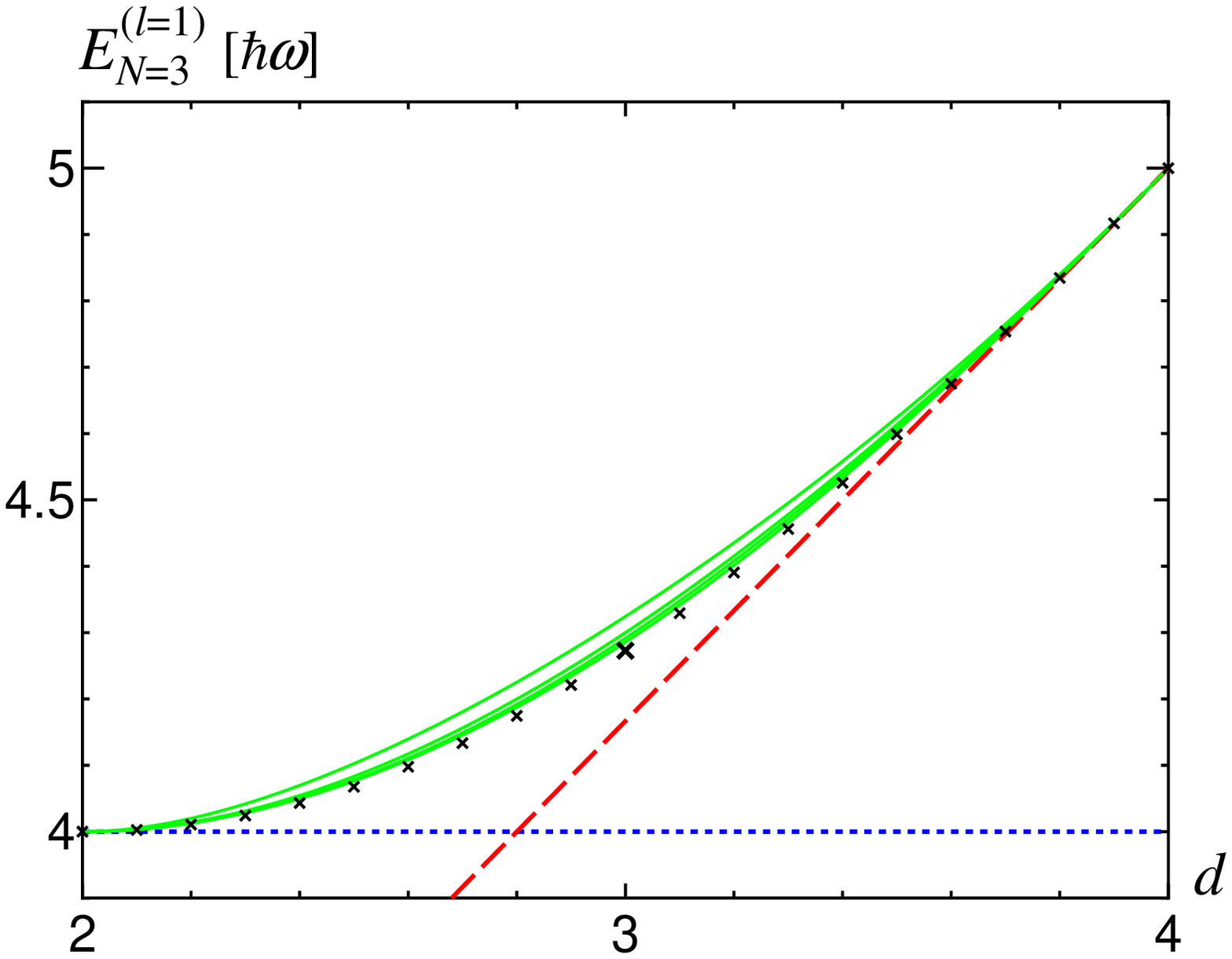}
 \caption{Energies of three fermions in a harmonic potential in the
 $s$-wave channel $l=0$ (left panel) and in the $p$-wave channel $l=1$
 (right panel) as functions of the spatial dimension $d$.  The dashed
 (dotted) lines are the extrapolations from the NLO expansions around
 $d=4$ in Eqs.~(\ref{eq:3-fermions_s-wave_4d}) and
 (\ref{eq:3-fermions_p-wave_4d}) [$d=2$ in
 Eqs.~(\ref{eq:3-fermions_s-wave_2d}) and
 (\ref{eq:3-fermions_p-wave_2d})].  The four solid curves show the
 Pad\'e interpolations of the two NLO expansions.  The symbols
 ($\times$) indicate the exact values for each $d$ obtained from
 Eqs.~(\ref{eq:anomalous}) and (\ref{eq:scaling}).
 \label{fig:3-fermion_energy}}
\end{figure}

Fig.~\ref{fig:3-fermion_energy} shows the behaviors of the three-fermion
energies $E_{N=3}^{(l)}$ with orbital angular momentum $l=0$ (left
panel) and $l=1$ (right panel) as functions of $d$.  The middle four
curves are the Pad\'e interpolations of the two NLO expansions.  Because
the exact results for arbitrary $d$ can be obtained from
Eqs.~(\ref{eq:anomalous}) and (\ref{eq:scaling}), we can use this case
as a benchmark test of our interpolation scheme.  We find that the
behaviors of the interpolated curves are quite consistent with the exact
results even within the leading corrections in $\eps$ and $\bar\eps$.
In $d=3$, these interpolations give
\begin{equation}
 \frac{E_{N=3}^{(l=0)}}{\omega} \approx 4.71,\ 4.7,\ 4.72,\ 4.72
\end{equation}
and
\begin{equation}
 \frac{E_{N=3}^{(l=1)}}{\omega} \approx 4.29,\ 4.3,\ 4.32,\ 4.29,
\end{equation}
which span very small intervals $E_3^{(0)}/\omega\approx4.71\pm0.01$
and $E_3^{(1)}/\omega\approx4.30\pm0.02$.  Our interpolated values are
reasonably close to the exact results $4.66622\,\omega$ and
$4.27272\,\omega$ in $d=3$~\cite{Tan:2004}.

Here we comment on the convergence of the $\eps$ expansions around $d=4$
and $d=2$.  By performing the expansions up to $O(\eps^{50})$ with the
use of the exact formula in Eq.~(\ref{eq:anomalous}) and studying their
asymptotic behaviors, we can find that the $\eps$ expansions are
convergent at least for these three-body problems.  Their radii of
convergence are estimated to be $|\eps|\lesssim0.48$ and
$|\bar\eps|\lesssim1.0$ for the $l=0$ case and $|\eps|\lesssim1.4$ and
$|\bar\eps|\lesssim1.0$ for the $l=1$ case.

The same analysis can be done for the energies of more than three
fermions in a harmonic potential where exact results are not available.
The Pad\'e interpolations of the two NLO expansions for $N=5$ yield
\begin{equation}
 \frac{E_{N=5}^{(l=0)}}{\omega} \approx 7.71,\ 7.64,\ 7.66,\ 7.82
\end{equation}
and
\begin{equation}
 \frac{E_{N=5}^{(l=1)}}{\omega} \approx 7.10,\ 7.16,\ 7.19,\ 7.09
\end{equation}
in $d=3$, which span relatively small intervals
$E_5^{(0)}/\omega\approx7.73\pm0.09$ and
$E_5^{(1)}/\omega\approx7.14\pm0.05$.  On the other hand, the Pad\'e
interpolations of the two NLO expansions for $N=4,\,6$ yield
\begin{equation}
 \frac{E_{N=4}^{(l=0)}}{\omega}
  \approx 5.55,\ 4.94,\ 4.94,\ 4.90,\ 6.17
\end{equation}
and
\begin{equation}
 \frac{E_{N=6}^{(l=0)}}{\omega}
  \approx 10.1,\ 7.92,\ 7.92,\ 7.80,\ 16.4
\end{equation} 
in $d=3$.  The first and last values corresponding to the Pad\'e
approximants where all terms are distributed to the numerator or
denominator are considerably off from the other three values.  This
would be because of the large NLO corrections near $d=4$.  If such two
extreme cases are excluded, the other three values span rather small
intervals $E_4^{(0)}/\omega\approx4.92\pm0.02$ and
$E_6^{(0)}/\omega\approx7.86\pm0.06$.  For comparison, the numerical
results obtained by using a basis set expansion technique are shown in
Table~\ref{tab:operator_2d}~\cite{Blume:2007}.

\section{General coordinate and conformal invariance \label{sec:general}}
Some nontrivial results can be obtained for the unitary Fermi gas using
general symmetry arguments.  For this end it is convenient to couple the
unitary Fermi gas to an external gauge field $A_\mu$ ($\mu=0,1,2,3$) and
to an external metric $g_{ij}$.  Both $A_\mu$ and $g_{ij}$ can be
functions of time and space.  Now the action of the unitary Fermi gas
becomes
\begin{equation}
 S = \int\!dt\,d\x \sqrt{g}\left( \frac i2 \psi_\sigma^\+\ D_t \psi_\sigma
   - \frac i2 D_t\psi_\sigma^\+\psi_\sigma
   - \frac1{2m} g^{ij}D_i\psi_\sigma^\+ D_j\psi_\sigma
   + \psi_\up^\+\psi_\down^\+\phi + \phi^\+\psi_\down\psi_\up \right),
\end{equation}
where $D_t=\d_t - iA_0$ and $D_i = \d_i - iA_i$ are covariant
derivatives.  We can see that $A_0$ plays the role of the external
trapping potential.  Recall that when the dimensional regularization is
used, the term $\propto\phi^\+\phi$ is absent in the unitarity limit.

By direct calculations, one can verify that this action is invariant
under the following transformations:
\begin{itemize}
\item Gauge transforms
\begin{subequations}
 \begin{eqnarray}
  && \psi \to e^{i\alpha(t,\x)}\psi, \qquad
   \phi \to e^{2i\alpha(t,\x)}\phi \\
  && A_0\to A_0-\d_t\alpha, \qquad A_i \to A_i -\d_i \alpha
 \end{eqnarray}
\end{subequations}
\item General coordinate transformations
\begin{subequations}
 \begin{eqnarray}
  && x^i \to x^{i'}, \qquad x^i = x^i(t,x^{i'}) \\
  && \psi(t,\x) \to \psi'(t,\x') = \psi(t,\x) \\
  && \phi(t,\x) \to \phi'(t,\x') = \phi(t,\x) \\
  && g_{ij}(t,\x) \to g_{i'j'}(t,\x')
   = \frac{\d x^i}{\d x^{i'}} \frac{\d x^j}{\d x^{j'}} g_{ij}(t,\x) \\
  && A_0(t,\x) \to A_{0'}(t,\x') = A_0(t,\x) + \dot x^i A_i 
   - \frac12 \dot x^i \dot x^j g_{ij} \\
  && A_i(t,\x) \to A_{i'}(t,\x') = \frac{\d x^i}{\d x^{i'}} A_i(t,\x)
   - \frac{\d x^i}{\d x^{i'}} \dot x^j g_{ij}(t,\x)
 \end{eqnarray}
\end{subequations}
\item Conformal transformations
\begin{subequations}
 \begin{eqnarray}
  && t \to t', \qquad t = t(t') \\
  && \psi(t,\x) \to \psi'(t',\x) = 
   \left(\frac{\d t}{\d t'}\right)^{3/4} \psi(t,\x) \\
  && \phi(t,\x) \to \phi'(t',\x) = 
   \left(\frac{\d t}{\d t'}\right) \phi(t,\x) \\
  && A_0(t,\x) \to A_0'(t',\x) = \left(\frac{\d t}{\d t'}\right) A_0(t,\x) \\
  && A_i (t,\x) \to A_i'(t',\x) = A_i(t,\x) \\
  && g_{ij}(t,\x) \to g'_{ij}(t',\x)
   = \left(\frac{\d t}{\d t'}\right)^{-1} g_{ij}(t',\x).
 \end{eqnarray}
\end{subequations}
\end{itemize}

These symmetries allow one to transform the unitary Fermi gas in a free
space into that in a harmonic potential with an arbitrary time-dependent
frequency $\omega(t)$.  This is done by a combination of a conformal
transformation $t=f(t')$, a general coordinate transformation
\begin{equation}
 x^i = \lambda^{-1}(t) x^{i'} \qquad\text{with}\qquad
  \lambda(t) = [f'(t)]^{-1/2},
\end{equation}
and a gauge transformation with
\begin{equation}
 \alpha =  \frac12 \frac{\dot\lambda}\lambda x^2.
\end{equation}
If one starts with $A_\mu=0$ and $g_{ij}=\delta_{ij}$, these three
transformations leave the gauge vector potential $A_i$ and the metric
$g_{ij}$ unchanged, but generate a scalar potential $A_0$:
\begin{equation}
 A_0 = \frac12 \omega^2(t)x^2 \qquad\text{with}\qquad
  \omega^2(t) = - \frac12 \frac{\ddot\lambda}\lambda x^2.
\end{equation}
The transformed field operator is
\begin{equation}
 \psi'(t,\x) = \exp\left( \frac i2 \frac{\dot\lambda}\lambda x^2\right)
  \lambda^{-3/2}(t) \psi(f(t), \lambda^{-1}(t)\x).
\end{equation}
This map between the unitary Fermi gas in the free space and that in the
harmonic potential with the time-dependent frequency was previously
found in Ref.~\cite{Castin:2004}.

In future applications, we only need the infinitesimal forms of the
transformations.  For reference, they are
\begin{subequations}\label{nonrel-gci}
 \begin{eqnarray}
  \delta\psi &=& i\alpha\psi -\xi^k\d_k\psi, \\
  \delta A_0 &=& -\dot\alpha-\xi^k\d_k A_0 - A_k \dot\xi^k,
  \label{nonrel-gci-A0} \\
  \delta A_i &=& -\d_i\alpha-\xi^k\d_k A_i - A_k\d_i\xi^k + mg_{ik}\dot\xi^k,
  \label{nonrel-gci-Ai} \\
  \delta g_{ij} &=& -\xi^k\d_k g_{ij} - g_{ik}\d_j\xi^k - g_{kj}\d_i\xi^k
 \end{eqnarray}
\end{subequations}
for the gauge and general coordinate transformations, and
\begin{equation}\label{conformal}
 \delta O = -\beta \dot O - \frac12\Delta[O] \dot\beta O
\end{equation}
for the conformal transformations, where $\Delta[O]$ is the dimension of
a field $O$; $\Delta[\psi]=\frac32$, $\Delta[\phi]=2$, $\Delta[A_0]=2$,
$\Delta[A_i]=0$, and $\delta[g_{ij}]=-2$.

\subsection{Vanishing bulk viscosities \label{sec:viscosity}}
One consequence of the general coordinate and conformal invariance is
the vanishing of the bulk viscosity of the unitarity Fermi gas in the
normal phase, and the vanishing of two (out of three) bulk viscosities
in the superfluid phase.  These conclusions come from the requirement
that hydrodynamic equations describing the motion of a fluid in the
external gauge field and metric possess the same set of symmetries as
the microscopic theory.

\subsubsection{Normal phase}
In the normal phase, the hydrodynamic equations are written in term of
the local mass density $\rho$, the local velocity $v^i$, and the local
entropy per unit mass $s$.  These equations are
\begin{eqnarray}
 && \frac1{\sqrt g}\d_t (\sqrt g\,\rho) + \nabla_i (\rho v^i) = 0,
 \label{continuity} \\
 && \frac1{\sqrt g}\d_t (\sqrt g\, \rho v_i) + \nabla_k \Pi^k_i 
 = \frac\rho m (E_i - F_{ik} v^k),
 \label{momentum-conservation} \\
 && \frac1{\sqrt g}\d_t(\sqrt g\,\rho s) 
 + \nabla_i\Bigl(\rho v^i\d_i s - \frac\kappa T\d^i T\Bigr) = \frac{2R}T,
 \label{S-prod}
\end{eqnarray}
where $\Pi_{ik}$ is the stress tensor, $\kappa$ is the thermal
conductivity, and $R$ is the dissipative function.  Compared to the
usual equations written for the flat metric and in the absence of the
gauge field, we have replaced the derivatives $\d_i$ by the covariant
derivatives $\nabla_i$ and added the force term in the momentum
conservation equation~(\ref{momentum-conservation}), which comes from
the electric force ($E_i=\d_t A_i-\d_i A_0$) and the magnetic Lorentz
force ($F_{ik}=\d_i A_k-\d_k A_i$). The stress tensor can be written as
\begin{equation}
 \Pi_{ik} = \rho v_i v_k + p g_{ik} - \sigma'_{ik},
\end{equation}
where $p$ is the pressure and $\sigma'_{ik}$ is the viscous stress
tensor.  The information about the kinetic coefficients is contained in
$\sigma'_{ik}$ and $R$.

In the dissipationless limit ($\sigma'=R=0$), the hydrodynamic equations
are invariant with respect to the general coordinate transformations,
provided that $\rho$, $s$, and $v^i$ transform as
\begin{eqnarray}
 \delta\rho &=& -\xi^k\d_k\rho, \\
 \delta s &=& -\xi^k\d_k s, \\
 \delta v^i &=& -\xi^k\d_k v^i + v^k\d_k\xi^i + \dot \xi^i.
\end{eqnarray}

Now consider the dissipative terms.  To keep the equations consistent
with the diffeomorphism invariance, one must require that $\sigma'_{ij}$
and $R$ transform as a two-index tensor and a scalar, respectively:
\begin{eqnarray}
 \delta\sigma'_{ij} &=& -\xi^k\d_k\sigma'_{ij} - \sigma_{kj}\d_i\xi^k
 -\sigma_{ik}\d_j\xi^k, \label{deltasigma} \\
 \delta R &=& -\xi^k \d_k R.
\end{eqnarray}
In a flat space the viscous stress tensor is given by
\begin{equation}
 \sigma'_{ij} = \eta(\d_i v_j + \d_j v_i) + \left(\zeta-\tfrac23\eta\right)
  \delta_{ij} \d_k v^k,
\end{equation}
where $\eta$ and $\zeta$ are the shear and bulk viscosities.  In the
naive extension to the curved space where one simply covariantizes the
spatial derivatives, $\sigma'_{ij}$ is not a pure two-index tensor; its
variation under the diffeomorphism contains extra terms proportional to
$\dot\xi^k$.  The correct extension is
\begin{equation}
 \sigma'_{ij} = \eta(\nabla_i v_j + \nabla_j v_i +\dot g_{ij})
  +\left(\zeta - \frac23 \eta\right) g_{ij} \left(\nabla_k v^k
    + \frac{\dot g}{2g} \right).
\end{equation}
Similarly, the dissipative function $R$ in the curved space becomes
\begin{multline}
 2R =
 \frac\eta 2 \left(\nabla_i v_j + \nabla_j v_i 
 - \frac23 g_{ij}\nabla_k v^k + \dot g_{ij} 
 - \frac13 g_{ij}\frac{\dot g}g \right)^2\\
 + \zeta \left(\nabla_i v^i + \frac{\dot g}{2g}\right)^2
 + \frac\kappa T\d_i T\d^i T.
\end{multline}

We then turn to the conformal invariance.  The dissipationless
hydrodynamic equations are invariant under~(\ref{conformal}), if the
dimensions of different fields are
\begin{equation}
 \Delta[\rho] = 2\Delta[\psi]= 3, 
  \qquad \Delta[s]=0, \qquad \Delta[v^i]=2.
\end{equation}
Now let us consider the dissipation terms.  From dimensional analysis,
one finds that one has to set
\begin{equation}
 \Delta[\eta] = \Delta[\zeta] = \Delta[\kappa] = \tfrac32
\end{equation}
for the hydrodynamic equations to be scale invariant.  However, the
conformal invariance is not preserved generically.  The culprit is
$\dot g_{ij}$ that transforms as
\begin{equation}
 \delta \dot g_{ij} = -\beta\ddot g_{ij} + \ddot\beta g_{ij},
\end{equation}
which leads to $\sigma'_{ij}$ and $R$ not to conform to the pattern
of~(\ref{conformal}), unless the bulk viscosity $\zeta$ vanishes.  Thus
the requirement of the conformal invariance of the hydrodynamic
equations implies $\zeta=0$.

\subsubsection{Superfluid phase}
Similarly, we can repeat the argument for the superfluid case.  The
hydrodynamics of superfluids contains an additional degree of freedom,
which is the condensate phase $\theta$, whose gauge-covariant gradient
is the superfluid velocity:
\begin{equation}\label{vs}
 v^s_i = \frac\hbar m (\d_i\theta + A_i).
\end{equation}
It transforms in the same way as the normal velocity $v_i \equiv v^n_i$
under the general coordinate and conformal transformations.  Its
consequence is that the relative velocity between the superfluid and
normal components $w^i=v_s^i-v^i$ transforms as a pure vector under the
diffeomorphism:
\begin{equation}
 \delta w^i = -\xi^k\d_k w^i + w^k\d_k \xi^i.
\end{equation}
The $\dot\xi^i$ term in the variation cancels between $\delta v_s$
and $\delta v$.

The diffeomorphism-invariant dissipative function in the curved space is
\begin{multline}
 2R =  \frac\eta 2 \left(\nabla_i v_j + \nabla_j v_i 
 - \frac23 g_{ij}\nabla_k v^k + \dot g_{ij} 
 - \frac13 g_{ij}\frac{\dot g}g \right)^2\\
 + 2\zeta_1 \left(\nabla_i v^i + \frac{\dot g}{2g}\right) \nabla_j (\rho_s w^j)
 + \zeta_2 \left(\nabla_i v^i + \frac{\dot g}{2g}\right)^2\\
 + \zeta_3 [\nabla_i(\rho_s w^i)]^2
 + \frac\kappa T \d_i T\d^i T.
\end{multline}
Under the conformal transformations, $R$ transforms as
\begin{equation}
 \delta R = -\beta\dot R - \frac72 \dot\beta R
  +\frac32\zeta_1\ddot\beta \nabla_i(\rho w^i)
  +\frac32\zeta_2\ddot\beta \left( \nabla_i v^i + \frac{\dot g}{2g}\right).
\end{equation}
The requirement of the conformal invariance of the superfluid
hydrodynamics implies that the $\ddot\beta$ terms must have vanishing
coefficients, i.e., $\zeta_1=\zeta_2=0$.

In conclusion, we find that in the unitary limit, the bulk viscosity
vanishes in the normal phase.  In the superfluid phase, two of the three
bulk viscosities vanish.

\subsection{Superfluid effective field theory \label{sec:EFT}}
At zero temperature, the long-distance dynamics of the unitary Fermi gas
is described by an effective field theory, with some effective action
$S_\mathrm{eff}$.  The effective theory should inherit the general
coordinate invariance of the microscopic theory.  This means that the
effective action is invariant under the general coordinate
transformations, which in its turn means that
\begin{equation}\label{deltaLscalar}
 \delta \mathcal{L} = -\xi^k \d_k \mathcal{L}.
\end{equation}
The low-energy degree of freedom is the phase of the condensate
$\theta$.  The time reversal symmetry means that the effective theory is
invariant under
\begin{equation}
 t\to -t \qquad\text{and}\qquad \theta \to -\theta.
\end{equation}

Since $\theta$ is a Nambu-Goldstone field, it should always appear with
derivatives in the effective Lagrangian.  Therefore, for power counting
purposes, we can set $\dot\theta, \d_i\theta\sim O(p^0)$.  The
leading-order effective Lagrangian should be
\begin{equation}
 \mathcal{L} = \mathcal{L} (\dot\theta, \d_i\theta, A_0, A_i, g_{ij}).
\end{equation}
The gauge invariance and the invariance with respect to
three-dimensional general coordinate transformations (with
time-independent $\xi^i$) limit the Lagrangian to be a function of two
variables:
\begin{equation}
 \mathcal{L} = \mathcal{L} \left( D_t\theta,\, g^{ij} D_i\theta D_j\theta\right),
\end{equation}
where
\begin{equation}
  D_t\theta = \dot\theta - A_0 \qquad\text{and}\qquad
  D_i\theta = \d_i\theta - A_i.
\end{equation}

The invariance of the effective theory with respect to general
coordinate transformations requires
\begin{equation}\label{LX}
 \mathcal{L} = P(X) \qquad\text{with}\qquad
  X = D_t \theta - \frac{g^{ij}}{2m} D_i\theta D_j\theta.
\end{equation}
If we now set the metric to be flat and the external field to be zero,
we find the most general form of the Lagrangian for superfluids:
\begin{equation}\label{GWW-L}
 \mathcal{L} = P \left(\dot\theta - \frac{(\d_i\theta)^2}{2m}\right),
\end{equation}
which was previously found by Greiter, Wilczek, and
Witten~\cite{Greiter:1989qb} using a different line of arguments.
Moreover, by studying the thermodynamics of the effective theory one
finds that the function $P(X)$ is the same function that determines the
dependence of the pressure on the chemical potential.

To the next-to-leading order, the effective Lagrangian contains terms
with two additional derivatives.  The symmetries restrict the number of
independent terms in the Lagrangian to two.  This allows one to relate
different physical quantities with each other~\cite{Son:2005rv}.

\section{Other scale and conformal invariant systems \label{sec:other}}
We have discussed various theoretical aspects of the unitary Fermi gas.
Finally we conclude this chapter by introducing other systems exhibiting
the nonrelativistic scaling and conformal symmetries, to which a part of
above results can be applied.  One such system is a mass-imbalanced
Fermi gas with both two-body and three-body resonances, which is already
mentioned in Sect.~\ref{sec:3-body}.  Its many-body physics is studied
in Ref.~\cite{Nishida:2007mr}.

\begin{table}[tp]
 \renewcommand\arraystretch{1.2}
 \caption{Eight classes of scale invariant nonrelativistic systems
 proposed in Ref.~\cite{Nishida:2008kr}.  In all cases below, the
 coupling of the interspecies contact interaction term has the dimension
 $[c_0]=-1$ and can be tuned to the nontrivial fixed point describing an
 interspecies resonance. \label{tab:mixed_D}}
 \begin{center}
  \begin{tabular*}{\textwidth}{@{\extracolsep{\fill}}lll}
   \hline\hline
   \ \# of species and dimensions & Spatial configurations
   & Symmetries $+\ M,\,D,\,C,\,H$ \ \\\hline
   \ 2 species in pure 3D & $\x_A=\x_B=(x,y,x)$
   & $P_i,\ K_i,\ J_{ij}\text{\ \ with\ \ }i,j=x,y,z$ \\
   \ 2 species in 2D-3D mixture & $\x_A=(x,y)\ \ \x_B=(x,y,z)$
   & $P_i,\ K_i,\ J_{ij}\text{\ \ with\ \ }i,j=x,y$ \\
   \ 2 species in 1D-3D mixture
   & $\x_A=(z)\ \ \x_B=(x,y,z)$ & $P_z,\ K_z,\ J_{xy}$ \\
   \ 2 species in 2D-2D mixture
   & $\x_A=(x,z)\ \ \x_B=(y,z)$ & $P_z,\ K_z$ \\
   \ 2 species in 1D-2D mixture
   & $\x_A=(z)\ \ \x_B=(x,y)$ & $J_{xy}$ \\
   \ 3 species in 1D-1D-1D mixture
   & $\x_A=(x)\ \ \x_B=(y)\ \ \x_C=(z)$ & None \\
   \ 3 species in 1D$^2$-2D mixture
   & $\x_A=\x_B=(x)\ \ \x_C=(x,y)$ & $P_x,\ K_x$ \\
   \ 4 species in pure 1D
   & $\x_A=\x_B=\x_C=\x_D=(x)$ & $P_x,\ K_x$ \\
   \hline\hline
  \end{tabular*}
 \end{center}
\end{table}

The other systems are multi-species Fermi gases in mixed
dimensions~\cite{Nishida:2008kr}.  In all systems listed in
Table~\ref{tab:mixed_D}, the coupling of the contact interaction term
that involves all species has the dimension $[c_0]=-1$ and thus the
theory with such an interaction term is renormalizable.  In particular,
when the coupling $c_0$ is tuned to its nontrivial fixed point
($c_0\neq0$) that describes an interspecies resonance, the resulting
system becomes scale invariant.  We can derive the reduced Schr\"odinger
algebra and the operator-state correspondence for such a
system~\cite{Nishida:2008gk}.  The few-body and many-body physics of
multi-species Fermi gases in mixed dimensions are studied in
Refs.~\cite{Nishida:2008kr,Nishida:2008gk,Nishida:2009fs,Nishida:2009nc,Nishida:2009pg}.
Some of these systems can be in principle realized in ultracold atom
experiments.  Indeed the 2D-3D mixture has been recently realized using
$^{41}$K and $^{87}$Rb and the interspecies scattering resonances were
observed~\cite{Lamporesi:2010}.


\section{Appendix: scaling dimensions of three-body operators}
\begin{figure}[t]
 \begin{center}
  \includegraphics[width=0.8\textwidth,clip]{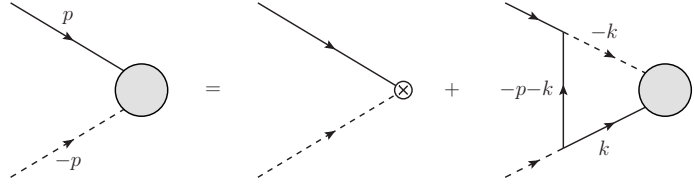}
 \end{center} 
 \caption{Feynman diagrams to renormalize three-body composite
 operators.  The solid lines are the propagators of $\psi_\up$ and
 $\psi_\down$ fields while the dotted lines are the resumed propagators
 of $\phi$ field.  The shaded bulbs represent the vertex function
 $Z(p)$. \label{fig:3-body_vertex}}
\end{figure}

In this Appendix, we derive the formula to compute the scaling
dimensions of three-body composite operators for arbitrary mass ratio
$m_\up/m_\down$, angular momentum $l$, and spatial dimension $d$ from
a field theory perspective.  We first consider a three-body operator
composed of two spin-$\up$ and one spin-$\down$ fermions with zero
orbital angular momentum $l=0$:
\begin{equation}
 \O_{\up\up\down}^{(l=0)}(\x) \equiv Z_\Lambda^{-1}\phi\psi_\up(\x),
\end{equation}
where $Z_\Lambda$ is a cutoff-dependent renormalization factor.  We
study the renormalization of the composite operator $\phi\psi_\up$ by
evaluating its matrix element $\<0|\phi\psi_\up(\x)|p,-p\>$.  Feynman
diagrams to renormalize $\phi\psi_\up$ is depicted in
Fig.~\ref{fig:3-body_vertex}.  The vertex function $Z(p_0,\p)$ in
Fig.~\ref{fig:3-body_vertex} satisfies the following integral equation:
\begin{eqnarray}\label{eq:integral}
 Z(p_0,\p) &=& 1 - i\int\!\frac{dk_0d\k}{(2\pi)^{d+1}}
  G_\up(k)G_\down(-p-k)D(-k)Z(k_0,\k) \nonumber\\
 &=& 1 - \int\!\frac{d\k}{(2\pi)^d}
  \left.G_\down(-p-k)D(-k)Z(k_0,\k)\right|_{k_0=\frac{\k^2}{2m_\up}},
\end{eqnarray}
where we used the analyticity of $Z(k_0,\k)$ on the lower half plane of
$k_0$.
$G_\sigma(p)\equiv\left(p_0-\frac{\p^2}{2m_\sigma}+0^+\right)^{-1}$ is
the fermion propagator and $D(p)$ is the resumed propagator of $\phi$
field given in Eq.~(\ref{eq:phi_propagator}).  If we set
$p_0=\frac{\p^2}{2m_\up}$, Eq.~(\ref{eq:integral}) reduces to the
integral equation for
$z(\p)\equiv Z\!\left(\frac{\p^2}{2m_\up},\p\right)$.

Because of the scale and rotational invariance of the system, we can
assume the form of $z(\p)$ to be
$z(\p)\propto\left(\frac{|\p|}\Lambda\right)^{\gamma}$, where $\Lambda$
is a momentum cutoff.  Accordingly the renormalization factor becomes
$Z_\Lambda\propto\Lambda^{-\gamma}$ with
$\gamma=-\d\ln Z_\Lambda/\d\ln\Lambda$ being the anomalous dimension of
the composite operator $\phi\psi_\up$.  In terms of $\gamma$, the
scaling dimension of the renormalized operator
$\O_{\up\up\down}^{(l=0)}$ is given by
\begin{equation}
 \Delta_{\up\up\down}^{(l=0)}
  = \Delta_{\phi}+\Delta_{\psi_\up}+\gamma = 2+\frac{d}2+\gamma.
\end{equation}
Substituting the expression of $z(\p)$ into Eq.~(\ref{eq:integral}) and
performing the integration over $|\k|$ at $\Lambda\to\infty$, we obtain
the following equation to determine $\gamma$:
\begin{equation}
 1 = \frac{2\pi^{1/2}
  \left[\frac{m_\down(2m_\up{+}m_\down)}{(m_\up{+}m_\down)^2}\right]^{1-d/2}}
  {\Gamma\!\left(1-\frac{d}2\right)
  \Gamma\!\left(\frac{d-1}2\right)\sin[(\gamma+1)\pi]}
  \int_0^\pi\!d\theta\,\sin^{d-2}\theta
  \frac{\sin[(\gamma+1)\chi]}{\sin\chi}
\end{equation}
with $\cos\chi\equiv\frac{m_\up}{m_\up+m_\down}\cos\theta$.  The
integration over $\theta$ can be done analytically in $d=3$, but
otherwise, has to be done numerically.

Similarly, for general orbital angular momentum $l$, we can derive the
equation satisfied by the anomalous dimension $\gamma_l$:
\begin{equation}\label{eq:anomalous}
 1 = \frac{2\pi^{1/2}
  \left[\frac{m_\down(2m_\up{+}m_\down)}{(m_\up{+}m_\down)^2}\right]^{1-d/2}}
  {\Gamma\!\left(1-\frac{d}2\right)
  \Gamma\!\left(\frac{d-1}2\right)\sin[(\gamma_l+l+1)\pi]}
  \int_0^\pi\!d\theta\,\sin^{d-2}\theta\,\tilde{P}_l(\cos\theta)
  \frac{\sin[(\gamma_l+l+1)\chi]}{\sin\chi},
\end{equation}
where $\tilde{P}_l(z)$ is a Legendre polynomial generalized to $d$
spatial dimensions.\footnote{$\tilde{P}_0(z)=1,\ \tilde{P}_1(z)=z,\ \ldots$.}  
The scaling dimension of the renormalized operator
$\O_{\up\up\down}^{(l)}$ with orbital angular momentum $l$ is now given
by
\begin{equation}\label{eq:scaling}
 \Delta_{\up\up\down}^{(l)}
  = \Delta_{\phi}+\Delta_{\psi_\up}+l+\gamma_l = 2+\frac{d}2+l+\gamma_l.
\end{equation}
$\Delta_{\up\up\down}^{(l)}$ for $l=0,1$ in $d=3$ are plotted as
functions of the mass ratio $m_\up/m_\down$ in
Fig.~\ref{fig:mass_ratio}, while $\Delta_{\up\up\down}^{(l)}$ for
$l=0,1$ with equal masses $m_\up=m_\down$ are plotted in
Fig.~\ref{fig:3-fermion_energy} as functions of the spatial dimension
$d$.

\end{document}